\documentclass[prl,twocolumn,superscriptaddress,longbibliography]{revtex4-1}
\usepackage{graphicx}
\usepackage{amssymb,amsmath,braket}
\usepackage{upgreek}
\usepackage{mathrsfs}
\usepackage{bm}
\usepackage[pagebackref=false,pdfnewwindow=true]{hyperref}
\usepackage[dvipsnames]{xcolor}
\usepackage{graphicx,tikz}
\usepackage[normalem]{ulem}
\usepackage{color, colortbl}
\usepackage{ulem} 

\usepackage{hhline}
\usepackage{array}
\usepackage{float}
\usepackage{multirow}

\DeclareMathOperator{\im}{Im}

\newcommand{\mac}{\mathcal}
\newcommand{\be}{\begin{equation}}
\newcommand{\ee}{\end{equation}}

\newcommand{\bit}{\begin{enumerate}}
\newcommand{\eit}{\end{enumerate}}

\newcommand*\diff{\mathop{}\!\mathrm{d}}
\newcommand{\non}{\nonumber}

\newcommand{\ve}[1]{{\boldsymbol #1}}

\newcommand{\thc}{\kappa^{\text{ph}}}
\newcommand{\trl}{\mathcal{T}}

\newcommand{\stkout}[1]{\ifmmode\text{\sout{\ensuremath{#1}}}\else\sout{#1}\fi}
\newcommand{\qv}{\ve{q}}
\newcommand{\kv}{\ve{k}}
\newcommand{\ver}{{\bf{r}}}

\newcommand*\Diff{\mathop{}\!\mathcal{D}}

\newcommand{\etaH}{\eta^H_{\Gamma\Gamma'}}
\newcommand{\GG}{{\Gamma\Gamma'}}

\newcommand{\pp}{{\scriptscriptstyle\perp}}
\newcommand{\scoc}{Sr$_2$CuO$_2$Cl$_2$}
\newcommand{\vhat}[1]{\bm{\hat{#1}}}

\def\em{\it}
\definecolor{bananayellow}{rgb}{1.0, 0.88, 0.21}
\definecolor{straw}{rgb}{0.32, 0.28, 0.1}

\allowdisplaybreaks

\begin{document}
\title{Phonon Hall Viscosity in Magnetic Insulators}
\author{Mengxing Ye}
\affiliation{Kavli Institute for Theoretical Physics, University of
  California, Santa Barbara, CA 93106, USA}
\author{Lucile Savary}
\affiliation{Universit\'e de Lyon, \'{E}cole Normale Sup\'{e}rieure de
  Lyon, Universit\'e Claude Bernard Lyon I, CNRS, Laboratoire de physique, 46, all\'{e}e
d'Italie, 69007 Lyon, France}
\author{Leon Balents}
\affiliation{Kavli Institute for Theoretical Physics, University of California, Santa Barbara, CA 93106, USA}
\date{\today}
\begin{abstract}
  The Phonon Hall Viscosity is the leading term evincing time-reversal
  symmetry breaking in the low energy description of lattice phonons.
  It may generate phonon Berry curvature, and can be observed
  experimentally through the acoustic Faraday effect and thermal Hall
  transport.  We present a systematic procedure to obtain the phonon
  Hall viscosity induced by phonon-magnon interactions in magnetic
  insulators under an external magnetic field. We obtain a general
  symmetry criterion that leads to non-zero Faraday rotation and Hall
  conductivity, and clarify the interplay between lattice symmetry,
  spin-orbit-coupling, external magnetic field and magnetic ordering.
  The symmetry analysis is verified through a microscopic
  calculation. By constructing the general symmetry-allowed effective
  action that describes the spin dynamics and spin-lattice coupling,
  and then integrating out the spin fluctuations, the leading order
  time-reversal breaking term in the phonon effective action, i.e.\
  the phonon Hall viscosity, can be obtained.   The analysis of the
  square lattice antiferromagnet for a cuprate Mott insulator,
  Sr$_2$CuO$_2$Cl$_2$, is presented explicitly, and the procedure
  described here can be readily generalized to other magnetic
  insulators.
\end{abstract}
\maketitle
{\em Introduction.---}  In a
material without time-reversal symmetry, the elastic stress may
contain a non-dissipative term proportional to the time derivative of
the elastic strain, the coefficient of which defines the phonon Hall
viscosity (PHV)~\cite{Barkeshli2012,HughesAFE2017,Shapourian2015}.
The PHV is analogous to the Hall viscosity in a fluid, which has been
heavily studied in quantum Hall states~\cite{Avron1995,Avron1998,Read2009,Read2011}, and
superconductors/superfluids\cite{Read2009,Read2011,Read2012}.
The PHV can be represented as a momentum space gauge field for
phonons, and as such encodes the Berry curvature of phonon
eigenstates.   Multiple mechanisms can
generate a PHV~\cite{AFETGG2010,Shi2012,Rosch2018,Xiao2019,Chen2020,Ye2020a}, including subtle band effects \cite{NagaosaPHE2019}.
In this paper, we consider the PHV generated by the magnetoelastic
coupling of strain and spins~\cite{AFETGG2010,Ye2020a}, taking into
account exchange and the Zeeman interaction of the spins with an external magnetic field.

The phonon Hall viscosity 
may be probed in experiments such as
acoustic Faraday rotation and thermal Hall transport. The first is a
measure of broken degeneracy between right and left circularly
polarized acoustic phonons;  the Faraday rotation angle is
proportional to the PHV~\cite{AFE1971,AFETGG2010,HughesAFE2017}.  The
second describes a transverse heat current in response to a thermal gradient, and the part of such heat current carried by phonons -- the
phonon Hall effect --  is proportional to the PHV (in the ballistic
phonon regime).   Recent thermal Hall
measurements in cuprate
compounds~\cite{Taillefer2019,Taillefer2020a,Taillefer2020b} suggest
that the phonon Hall effect may dominate the thermal Hall signal in
both the Mott insulating and pseudogap phases of cuprates.  

In this paper, we present a systematic procedure to obtain the PHV in
magnetic insulators, based on the coupling of spins to elastic
strains.  We concentrate mainly on the PHV terms that contribute to
the acoustic Faraday rotation and phonon Hall conductivity.  We first
introduce the effective field theory and the expression for the
viscosity coefficients.  We follow with a thorough symmetry analysis in
different phases and regimes, arriving at a symmetry criterion for a
nonzero PHV, which we verify through a microscopic calculation.  We
conclude with a discussion of experimental implications, spin-orbit
coupling, and open questions.

{\em Phonon effective action and magnetoelastic coupling---}The PHV term in the
effective linear response action is~\cite{Avron1998,Barkeshli2012},
\begin{align}
    \mathcal{S}_{\rm PHV}=\frac{1}{2}\sum_{\qv,\Gamma,\Gamma'} \int \diff t \,\, \etaH(\qv) (\mathcal{E}_{\Gamma,\qv} \dot{\mathcal{E}}_{\Gamma',-\qv}-\dot{\mathcal{E}}_{\Gamma,\qv} \mathcal{E}_{\Gamma',-\qv}),
    \label{eq:PHV}
\end{align}
where $\mathcal{E}_{ij}=\frac{1}{2}(\partial_i u_j+\partial_j u_i)$ is
the strain tensor, $i,j$ label Euclidean coordinates, and $\ve{u}$ is
the lattice displacement field. Here
$\dot{\mathcal{E}}=\partial_t \mathcal{E}$, and $\Gamma,\Gamma'$ are
the irreducible representations (irreps) of a group whose precise
definition we make clear below. The Hall viscosity
$\eta^H_{\Gamma\Gamma'}(\qv)$ describes the stress $\varSigma_\Gamma$
resulting from a time-dependent strain $\mathcal{E}_{\Gamma'}$, i.e.
$\varSigma_{\Gamma}\sim \delta\mathcal{S}_{\rm
  PHV}/\delta{\mathcal{E}_{\Gamma}}\sim \eta_{\Gamma
  \Gamma'}\dot{\mathcal{E}}_{\Gamma'}$~\cite{Read2011}. As we will
discuss at length below, symmetry constrains greatly which
$\Gamma\Gamma'$ lead to a nonvanishing $\eta^H_{\Gamma\Gamma'}$. We
allow for a momentum dependence of the Hall viscosity
$\eta^H_{\Gamma\Gamma'}(\qv)=-\eta^H_{\Gamma'\Gamma}(-\qv)$.\footnote{
  Symmetry allows additional PHV terms which are not expressed solely
  in terms of the strain but also in terms of the time derivative of
  the rotation
  $\mathcal{M}_{ij}=\frac{1}{2}(\partial_i u_j-\partial_j u_i)$. Such
  terms, however, cannot be induced by conventional magnetoelastic
  coupling which involves strain only.}

The Lagrangian density for the magnetoelastic coupling can be expressed as
\begin{align}
\mac{L}_{sl}=\sum_{\Gamma,a}\lambda_{\Gamma,a}\,\mathcal{E}_\Gamma(\ve{x},\tau)\cdot\mac{O}_{\Gamma,a}(\ve{x},\tau),
    \label{eq:Lsl}
\end{align}
where $\lambda_{\Gamma,a}$ denotes the magnetoelastic coupling strength
in units of energy to the $\Gamma$th component of the strain field
$\mathcal{E}_\Gamma$ of (each independent copy $a$ of) the spin operator
composite $\mac{O}_{\Gamma,a}(\ve{x},\tau)$ which transforms under the
same $\Gamma$ irrep.  

The Hall viscosity itself is then
obtained by extracting terms linear in the phonon energy
$\omega$ (reflecting the fact that the PHV is
time-reversal odd). More precisely, the Hall viscosity ``coefficient'' $\eta^H_{\Gamma\Gamma'}$ is obtained from the correlation of $\mac{O}_\Gamma$, $\mac{O}_{\Gamma'}$ as~\cite{Ye2020a}:
\begin{equation}
  \label{eq:1}
  \frac{i \eta^H_{\Gamma\Gamma'}(\qv)}{\lambda_\Gamma\lambda_{\Gamma'}}=\lim_{\omega\rightarrow
    0}\partial_\omega\sum_{a,a'}\langle \mac{O}_{\Gamma a}(-\qv,\tau) \cdot\mac{O}_{\Gamma'a'}(\qv,0)\rangle_{-i\omega+0^+},
\end{equation}
where $\langle \mathcal{A}\rangle_{\omega_n}\equiv\int_0^\beta \diff
\tau e^{i\omega_n \tau}\langle T_\tau\mathcal{A}(\tau)\rangle$ denotes the
Fourier transform, $\tau$
  is imaginary time, $\omega_n=2\pi n/\beta$ is the $n$th bosonic
  Matsubara frequency and $T_\tau$ denotes imaginary-time ordering.

  {\em Magnetic system.---}We now specify the theory to calculate the
  correlations of the $\mathcal{O}$ operators.  We study a two-dimensional spin system
  which orders antiferromagnetically in zero field.  Because we will
  compute $\etaH(\qv)$ for small momentum $\qv$ at low temperature
  (compared to the Debye temperature and magnon bandwidth), it is
  enough to consider low-energy spin and lattice
  fluctuations.    We therefore employ a path
  integral approach, and describe the low energy dynamics in terms of
  continuous fields: the irrep components of the strain tensor
  ($\mathcal{E}_\Gamma$) and the staggered ($\ve{n}$) and uniform
  ($\ve{m}$) magnetizations.  The latter two fields are described by
  the well-known nonlinear sigma model (see SM), with the constraints
  $|\bm{n}|^2=1$ and $\bm{n}\cdot\bm{m}=0$.

The $\mathcal{O}$ correlations are obtained from the nonlinear sigma
model as follows.  We will consider a weak applied field along the $z$-axis, $\bm{h}=h\bm{\hat{z}}$.  We assume that in the zero-field limit
the uniform magnetization vanishes,
$\lim_{h\rightarrow 0} \langle\bm{m}\rangle=0$, and the N\'eel vector
$\bm{n}$ lies along the $x$-axis,
$\lim_{h\rightarrow 0} \langle \bm{n}\rangle = n_0\bm{\hat{x}}$.  Then
constraints are solved via $\bm{n}=(n_0,\bm{n}_\pp)$, with
$n_0=\sqrt{1-\bm{n}_\pp^2}$, $\bm{m}=(m_x,\bm{m}_\pp)$ and
$m_x = -\bm{n}_\pp \cdot \bm{m}_\pp/n_0$.  We also define
${\bf{m}}$ through $\bm{m}=\chi h \bm{\hat{z}}+{\bf{m}}$,
highlighting the field-induced uniform magnetization
$\chi h \bm{\hat{z}}$ ($\chi$ is the magnetic susceptibility
  along the $z$-axis).  $\bm{n}_\pp,{\bf{m}}_\pp$ form two
canonically conjugate pairs of Gaussian fields and may be considered
``spin-wave variables''. We also define ${\upomega}_{\alpha,\bm{k}}$
to be the dispersion for the magnon branch $\alpha\in \{y,z\}$.  In
the case of a Heisenberg model with exchange $J$, plus weak anisotropies
of the order of $\Delta_\alpha$, the dispersion relation takes the form
$\upomega_{\alpha,\bm{k}}=v_m\sqrt{k_x^2+k_y^2+\delta_\alpha^2}$,
where $v_m$ is the magnon velocity and $\delta_{\alpha}$ are the
magnon gaps in units of inverse length:
$\delta_y=\Delta_y/v_m$,
$\delta_z=\sqrt{\Delta_z^2+h^2}/v_m$. 

From this, we find that correlators of the form $\langle
nn\rangle, \langle {\rm mm}\rangle$ are even functions of frequency and
given in the SM, and only the ``mixed'' correlators $\langle
n{\rm m}\rangle$ are odd in frequency:
 \begin{align}
    &\langle n_{\alpha,\bm{k}}
      {\rm m}_{\bar{\alpha},-\bm{k}}\rangle_{\omega_n} =
      \epsilon_{x\alpha\bar{\alpha}}\, \frac{\omega_n a_0^2}{S} \mac{D}_\alpha(\bm{k},\omega_n),
    \label{eq:correlator}
\end{align}
where $\epsilon$ is the Levi-Civita tensor,
$\overline{\alpha}\neq\alpha,x$, $a_0$ is the lattice constant of the magnetic layer, and 
$\mac{D}_\alpha^{-1}(\bm{k},\omega_n)=\omega_n^2+\upomega_{\alpha,\bm{k}}^2$.

The magnetic operators $\mathcal{O}_{\Gamma,a}$ may be expressed as
polynomials in the staggered and uniform magnetizations
$\bm{n},\bm{m}$.  In order to understand their form, we must proceed
to a thorough symmetry analysis.
  
\begin{table*}
\centering
\renewcommand{\arraystretch}{1.5}
\begin{tabular}{c|l|l|l}
\hline
&  \multirow{2}{3cm}{\quad\quad\quad\quad zero field} & \multicolumn{2}{c}{$\ve{h}=h\bm{\hat{z}}$}\\
\cline{3-4}
&   & \quad\quad lattice and spin & \quad lattice effective \\
\hline
 paramagnet & \;$\mathsf{G}=P4/mmm1'$\; &
                                          \;$\mathsf{G}(\bm{0},h\bm{\hat{z}})=P4/mm'm'$  &  \;$\mathsf{G}^{\rm eff}(\bm{0},h\bm{\hat{z}})=4/mm'm'$ \\
  \hline
  high sym.\ AFM & \;$\mathsf{G}(\bm{\hat{x}},\bm{0})=\langle
                   i,\trl X,\trl Y, C_{2x},  \trl C_{2z}\rangle$\;
                                                      &\;$\mathsf{G}(\bm{\hat{x}},h\bm{\hat{z}})=\langle
                                                        i,XY,
                                                         \trl C_{2y},
                                                        X C_{2z}
                                                        \rangle$ &
                                                                   \;$\mathsf{G}^{\rm
                                                                   eff}(\bm{\hat{x}},h\bm{\hat{z}})=\langle
                                                                   i ,
                                                                   \trl
                                                                   C_{2y}, C_{2z} \rangle$ \\
\hline
 low sym.\ AFM & \;$\mathsf{G}(\bm{\hat{e}},\bm{0})=\langle
                 i,\trl X,\trl Y, \trl C_{2z}\rangle$\; &
                                                         \;$\mathsf{G}(\bm{\hat{e}},h\bm{\hat{z}})=\langle i, XY, X C_{2z} \rangle$ & \;$\mathsf{G}^{\rm eff}(\bm{\hat{e}},h\bm{\hat{z}})=\langle i, C_{2z}\rangle$ \\
\hline
\end{tabular}
\caption{\label{tab:MagneticSG} Summary of the magnetic groups (in the
  Hermann-Mauguin notation) for crystallographic layer group
  $G=P4/mmm$. Rows denote different values for $\bm{n}$: without
  magnetic order, and for the two distinct orientations of the
  staggered magnetization in the $xy$ plane (here $\bm{\hat{e}}$ is a
  generic vector {\em not} along a high-symmetry axis in the plane), and
  columns specify the zero and finite field cases, and the effective
  magnetic group for the effective lattice theory.  Here $i$ is inversion, $P4/mmm1'=P4/mmm\times ({\rm Id}+ \mathcal{T})$ is a grey
  group,
  $P4/mm'm'=\left({\rm C}_{4h}+ \trl\times ({\rm D}_{4h}-{\rm C}_{4h})\right)\ltimes P$
  is a black-white group in which $m'$ denotes the vertical mirror
  composed with $\trl$.  The symbol $\langle \cdot \rangle$ indicates
  the group generated by the ``$\cdot$'' operations.}
\end{table*}

{\em Symmetry group considerations.---}  We call $G$ the lattice space
group of the crystal in the paramagnetic phase. We
also define $\mathsf{G}$ to be its associated (paramagnetic) {\em magnetic} space
group, $\mathsf{G}=G\times\{{\rm Id},\mathcal{T}\}$.  Spontaneous
ordering with staggered magnetization $\bm{n}$ and/or an external
field $\bm{h}$ reduce the system's symmetry to
$\mathsf{G}(\bm{n},\bm{h})$.  For example, the subgroup
of $\mathsf{G}$ which preserves the AFM order defines the magnetic
space group of the AFM, $\mathsf{G}(\bm{n},\bm{0})$, and the subgroup which
preserves the Zeeman interaction with the magnetic field $\bm{h}$ defines the magnetic space group of the paramagnet in a field, $\mathsf{G}(\bm{0},\bm{h})$.  In simple cases, the magnetic space group of the AFM
in an external field, $\mathsf{G}(\bm{n},\bm{h})$, can be found as the intersection set of
$\mathsf{G}(\bm{n},\bm{0})$ and
$\mathsf{G}(\bm{0},\bm{h})$.

Ultimately the PHV is defined in the {\em effective}
action, $\mathcal{S}_{\rm PHV}$ from
Eq.~\eqref{eq:PHV}, purely in the phonon space. It results, in the framework described here,
from integrating out the magnetic degrees of freedom $\bm{m},\bm{n}$,
i.e.\ from carrying out the integral in Eq.~(A2) of the
SM.  One may then observe a simplifying feature of the {\em representation} of
the symmetry operations on the lattice terms. In particular, {\em the
  translation operation acts as the identity on the lattice strain
  field}.  Indeed, in $\mathcal{L}_{sl}$, $\mathcal{O}_\Gamma$
and $\mathcal{E}_\Gamma$ are {\em independently} translationally
invariant. As a consequence, the appropriate representation of
$\mathsf{G}(\bm{n},\bm{h})$ acting on the terms in the effective
phonon action (in the presence of AFM order and in a field) is that of
the {\em magnetic point group} $\mathsf{G}^{\rm eff}(\bm{n},\bm{h})$ obtained
through the group morphism
$\Pi:(W,t)\mapsto(W,0)$, where $(W,t)\in \mathsf{G}(\bm{n},\bm{h})$
and $W,t$, are respectively the (anti)linear and translational parts
of $(W,t)$ and $0$ denotes here the zero translation. The
morphism theorem applies and 
\begin{equation}
  \label{eq:2}
  \mathsf{G}^{\rm eff}(\bm{n},\bm{h})\cong\mathsf{G}(\bm{n},\bm{h})/\mathsf{T}(\bm{n},\bm{h}),
\end{equation}
i.e.\ $\mathsf{G}^{\rm eff}(\bm{n},\bm{h})$ is isomorphic to the
factor group $\mathsf{G}(\bm{n},\bm{h})/\mathsf{T}(\bm{n},\bm{h})$, where $\mathsf{T}(\bm{n},\bm{h})$ is the
pure translation subgroup of $\mathsf{G}(\bm{n},\bm{h})$.
The group $\mathsf{G}^{\rm eff}(\bm{n},\bm{h})$ may now be
used to analyze the PHV.   In particular, distinct irreps
$\Gamma,\Gamma'$ under ${\sf G}$
may collapse to the same irrep under $\mathsf{G}^{\rm
  eff}(\bm{n},\bm{h})$, allowing the corresponding strain components $\mathcal{E}_\Gamma,\mathcal{E}_{\Gamma'}$
to couple in the PHV term.  

{\em Application.---} As an example, we now consider a 3D crystal
composed of finite thickness regions whose symmetry is given by the
{\em layer group} $G=P4/mmm$ (layer group number 61, see e.g.\
Ref.~\cite{Aroyo2006}), whose point group is ${\rm D}_{4h}$ in the
sense that $G$ is the semi-direct product of the point group $4/mmm$
(${\rm D}_{4h}$) and the 2D translation group $P$ of a square Bravais
lattice.  The layer group is the natural framework to describe those
properties of quasi-2d solids which are independent of the stacking
structure (however, some stackings may reduce the point group
symmetries, and thereby relax restrictions on the PHV).  The cuprate
Mott insulator Sr$_2$CuO$_2$Cl$_2$ with space group $I4/mmm$ (number
139) is such an example.  The translation group $P$ is generated by
translations $X,\, Y$ within the $xy$ plane, i.e.\ by the Bravais
lattice vectors $\vhat{x}, \vhat{y}$, respectively.  To characterize
the lattice strain field, only the \emph{point} group is relevant, and
the decomposition of the strain tensor $\mathcal{E}$ into irreducible
representations (irreps) of ${\rm D}_{4h}$ reads:
\begin{align}
    \mathcal{E}_{A_1}&= \mathcal{E}_{xx}+\mathcal{E}_{yy},\, \mathcal{E}_{B_1}=\mathcal{E}_{xx}-\mathcal{E}_{yy},\, \mathcal{E}_{B_2}=\mathcal{E}_{xy},\non\\ \mathcal{E}_{E}&=\{\mathcal{E}_{Ex},\mathcal{E}_{Ey}\}=\{\mathcal{E}_{xz},\mathcal{E}_{yz}\}.
    \label{eq:strainT}
\end{align}
In the quasi-2D limit, we ignore the
inter-layer spin-lattice coupling, so $\mathcal{E}_{zz}$ in the $A_1$
irrep does not enter the Hall viscosity.

Consider next the determination of $\mathsf{G}(\bm{n},\bm{h})$ and
$\mathsf{G}^{\rm eff}(\bm{n},\bm{h})$ for this system.  For concreteness, we choose
the field along the $z$-axis,  $\ve{h}=h\bm{\hat{z}}$, and place
$\bm{n}$ in the $xy$ plane (which is energetically preferred for this
field orientation).  For our purpose, we distinguish two cases:  the ``high symmetry''
AFMs for which $\bm{n}$ is aligned with a high-symmetry direction,
i.e.\ to the $x$ or $y$ axes or at 45 degrees between
them, and the ``low symmetry'' AFMs where $\bm{n}$ takes any other in-plane
orientation.    The paramagnetic and
magnetic groups for zero and nonzero fields with the above provisos
are given in Table~\ref{tab:MagneticSG}.

For this field orientation, we are interested in $\kappa_{xy}$ and the
acoustic Faraday effect for sound waves propagating along
$\bm{\hat{z}}$.  Both of these are odd under time-reversal and under
vertical mirrors, and hence even under their combination.  This is
compatible with all three cases of $\mathsf{G}^{\rm
  eff}(\bm{n},\bm{h})$ in a non-zero applied field (see the final column
of Table~\ref{tab:MagneticSG}), and hence both effects should be non-zero in
these situations.  To determine them, we will need the PHV terms with
the same symmetries, i.e.\ those which are odd under break time-reversal $\mathcal{T}$ and vertical mirrors, and
which are invariant under $\mathsf{G}^{\rm eff}(\bm{n},\bm{h})$: the only compatible PHV coefficients are
$\eta^H_{B_1,B_2}$ and $\eta^H_{E_x,E_y}$.  Other field
configurations can be analyzed similarly -- see
SM.

{\em Contributing magnetic operators.---}We are now in a position to ask about the allowed forms of the $\mathcal{O}$ operators
which appear in $\mathcal{L}_{sl}$ and which contribute to a nonzero
$\eta^H$, as well as their correlations. The magnetic operators
$\mathcal{O}$ are polynomials of the Gaussian fields
$\bm{n}_\pp,{\bf{m}}_\pp$, so that we may calculate correlators
of the $\mathcal{O}$'s
using Wick's theorem. We keep only
those which effectively yield ``mixed'' correlators of ${\rm m}$
and $n$.

In the absence of a magnetic field, the PHV vanishes. Indeed,
because $\bm{m}$ is odd under the $\mathcal{T}X$ and $\mathcal{T}Y$ symmetries of the AFM state,
all allowed polynomials are even in $\bm{m}$. In an external magnetic
field however, the
uniform magnetization $\ve{m}$ acquires a static component
$\ve{m}_0 =\chi h\bm{\hat{z}}$ so that
${\rm m}_x = -\chi h n_z$ to linear order in $h$.  
One can thereby obtain an odd-in-frequency contribution to the
two-point $\mathcal{O}$ correlators: this occurs in the cases for which one magnetic
operator $\mathcal{O}$ is
even and the other is odd in ${\bf m}_\pp$.  To lowest
order, the important terms are $\mathcal{O}_{B_1} \sim h {\rm m}_z$,
$\mathcal{O}_{B_2}\sim n_y$, $\mathcal{O}_{E_x}\sim n_z$,
$\mathcal{O}_{E_y} \sim h {\rm m}_y$ (see SM Sec.B3) ---contributions from higher powers of magnon operators are
parametrically small by a factor of $k_BT/J$ or $v_m \delta_\alpha/J$ (see
SM Sec.~D).  Appropriately combined, the latter give
\begin{align}
    \eta_{B_1,B_2}^H\sim -h \langle   {\rm m}_z n_y \rangle,\quad
    \eta_{E_x,E_y}^H &\sim h \langle   n_z {\rm m}_y  \rangle.
    \label{eq:eta1}
\end{align}
A careful analysis that restores the units (see SM Sec.~A1)
gives 
\begin{align}
    \etaH(\qv) = \frac{\lambda_\Gamma \lambda_{\Gamma'} S^3 \chi h
  }{d_za_0^2 } \mac{D}_{\alpha(\Gamma\Gamma')}(\qv_\perp,\omega=0)=  \frac{\gamma_{\Gamma\Gamma'}}{q_x^2+q_y^2+\delta_\alpha^2},
    \label{eq:eta2}
\end{align}
where $\gamma_{\Gamma\Gamma'}=\frac{hS^2\lambda_\Gamma \lambda_{\Gamma'}
  }{v^3_m g  d_z a_0}$, $\alpha(B_1,B_2)=y$, $\alpha(E_x,E_y)=z$, and
  $\qv_\perp$ projects the momentum $\qv$ to the xy plane.  The factor of
$d_z$, the inter-layer spacing in the $z=c$ direction, is necessary to
convert to bulk three-dimensional elasticity.

\emph{Acoustic Faraday effect.---}  This effect can be understood as
arising from splitting of the degeneracy between right- and
left-circularly polarized sound waves.  This allows a 
transverse linearly-polarized wave propagating along $\bm{\hat{z}}$ at a frequency
$\omega_{\rm ph}$ to undergo a Faraday rotation.  The Faraday rotation
angle $\Phi$ per unit length $L$ is simply given by the difference in
wavenumber between right- and left-circularly polarized waves at a
given frequency $\omega_{\rm ph}$.  By calculating the dispersion relation including the
PHV, we find $\Phi/L$ is related to the Hall viscosity
coefficient $\eta^H_{E_xE_y}$ through (see the SM)
\begin{align}
\frac{\Phi}{L}=\frac{\eta^H_{E_xE_y} \omega_{\rm
  ph}^2}{v_T^3\rho}=\frac{\gamma_{E_x E_y}\omega_{\rm
  ph}^2}{v_T^3\rho\delta_z^2}+{O}(\omega_{\rm ph}^3).
\label{eq:AFE}
\end{align}
where $\rho$ is the mass density of the lattice, $v_T$ is the
asymptotic long wavelength transverse sound wave velocity for this
propagation direction, and recall $\delta_z$ is the 
out-of-plane magnon gap in units of inverse length.


\begin{figure}[h!]
    \centering
    \includegraphics[width=0.9\columnwidth]{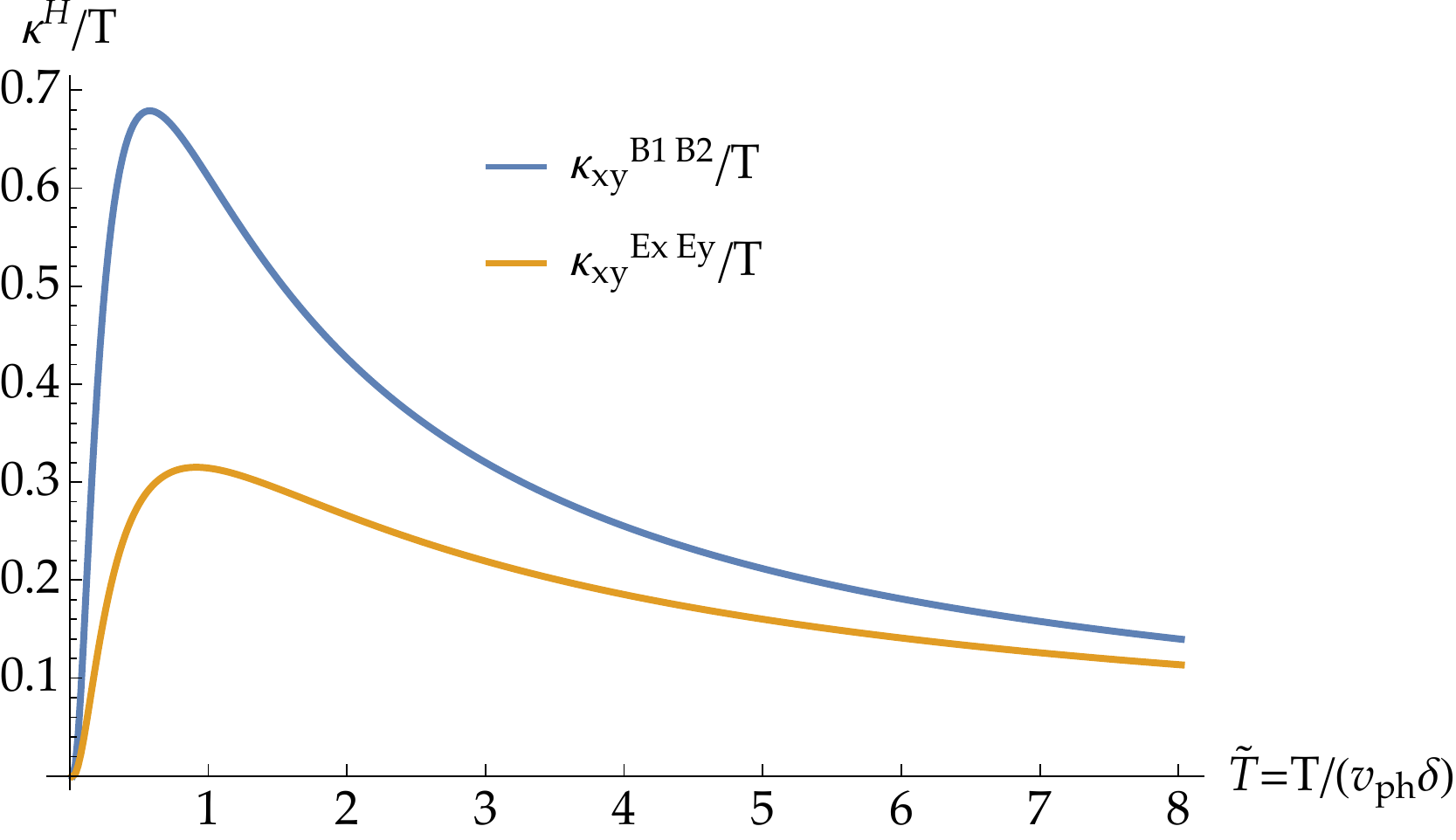}
    \caption{Intrinsic phonon thermal Hall conductivity,
      $\kappa_{xy}/T$ v.s.\ $T$ (rescaled in units of $v_{\rm ph}
      \delta_\alpha$). Only the value of $F_\alpha(\tilde T)$ in
      Eq.~\eqref{eq:kappaint} is plotted, and the overall coefficient
      not shown in the plot is set by the RHS of
      Eq.~\eqref{eq:kappaval}.}
    \label{fig:kappa}
\end{figure}

\emph{Thermal Hall conductivity.---}Following Ref.~\cite{Shi2012}, the
intrinsic (by which we mean independent of impurities) non-collisional thermal Hall
conductivity $\kappa_{jk}$ is determined by the Berry curvature
$\varOmega^i_{\qv,\sigma}$ and the phonon dispersions
$\upomega_{\qv,\sigma}$:
\begin{align}
\label{eq:3}
    \kappa_{jk}=-\frac{1}{T}\int_0^\infty \diff E\, E^2 \sigma_{jk}(E) \frac{\diff n^{\text{eq}}_B(E)}{\diff E},
\end{align}
where we defined the ``conductivity'' $\sigma_{jk}$:
\begin{align}
  \label{eq:4}
    \sigma_{jk}(E)=-\sum_\sigma \int \frac{\diff^3 q}{(2\pi)^3}\epsilon_{ijk}\varOmega^i_{\qv,\sigma}\Theta(E-\upomega_{\qv,\sigma}),
\end{align}
where $\epsilon$ is the Levi-Civita tensor, the subscript $\sigma$ labels the phonon branch, and $n_B^{\rm eq}$ is the
Bose-Einstein distribution.

A non-zero Berry curvature $\bm{\varOmega}$ is obtained by finding the
phonon eigenfunctions including the PHV, which induces a vector
potential in the phonon Hamiltonian (see Sec.~C2 of the
SM).   The leading nonzero terms are linear in
$\eta^H$, and to this order contributions to $\kappa_H$ from $\eta^H_{\Gamma\Gamma'}$
add--each is labeled $\kappa_H^\GG$ below. Defining a
dimensionless temperature $\tilde{T}=T/(v_{\rm ph}\delta_\alpha)$, where
$v_{\rm ph}$ is the mean sound velocity, we find
\begin{align}
\frac{\kappa^{\GG}_{xy}}{T}\sim \frac{\gamma_{\Gamma\Gamma'}}{\rho
  v_{\rm ph}}F_{\alpha(\Gamma\Gamma')}(\tilde{T}),
\label{eq:kappaint}
\end{align}
where $F_{\alpha(\Gamma\Gamma')}$ is a scaling function. 
The detailed forms of $F_{\alpha(\Gamma\Gamma')}(\tilde{T})$ for different $\Gamma\Gamma'$ are given in the SM, and are
numerically evaluated and plotted in Fig.~\ref{fig:kappa}.
Importantly, we find that the temperature scaling of $\kappa_{H}/T$ is sensitive to the
$\qv$ dependence of $\eta^H$. If $\eta^H$ is a constant, straightforward
power counting shows that $\kappa_H/T \sim T^{d-1}$, where $d$ is the
spatial dimension. For $\eta^H$ with the $\qv$ dependence of Eq.~\eqref{eq:eta2}, the spin gap $\sim\delta_\alpha$ introduces an
additional scale that controls the temperature dependence, and
$\kappa_H/T$ is non-monotonic. $\kappa_H/T$ increases from $T=0$ as $T^2$,
reaches a maximum at $T_{\rm max}\sim v_{\rm ph} \delta_\alpha$. At
$T/v_{\rm ph}\delta_\alpha\gg1$, $\kappa_H/T$ decreases as $1/T^\zeta$, where
$\zeta\sim 1$ (see Fig.~\ref{fig:kappa}).

\emph{Summary and discussion.---}In this work, we presented a symmetry
analysis of the phonon Hall viscosity and resulting phonon Berry
curvature in a magnetoelastically coupled system.  The procedure
includes accounting for separate scales for anisotropy-induced magnon
gaps and the applied magnetic field, and follows from a magnetic space
group symmetry analysis. The symmetry predictions were checked for a
low-energy magnon model by an explicit calculation in terms of spin
correlation functions.  We found that the phonon Hall viscosity
$\eta^H$ retains non-trivial dependence on the the scaling variable
$\qv\delta$, even when the phonon momentum is small $\qv a_0 \ll 1$,
due to the small spin gap $\sim \delta$.

As an example, we modeled the cuprate Mott insulator
Sr$_2$CuO$_2$Cl$_2$ with tetragonal symmetry. We showed that PHV
induces thermal Hall conductivity via both Hall viscosities
$\eta^H_{B_1 B_2}$ and $\eta^H_{E_x E_y}$, while the acoustic Faraday
effect arises only from $\eta^H_{E_x E_y}$.

The treatment above implicitly uses spin-orbit coupling (SOC)
throughout, in the symmetry analysis and through the forms of the
spin-lattice couplings.  Without SOC, one instead requires that the
microscopic Hamiltonian (before any spontaneous symmetry breaking) in
the absence of an applied magnetic field has a global SO(3)
spin-rotation symmetry $SO(3)_s$. The latter acts only on spin indices, and
is distinct from and independent of the space group and time-reversal
symmetries, which act on coordinates, spatial derivatives, and strain
indices.  Under the assumption that the applied magnetic field couples
only via the Zeeman interaction, then the symmetry constraints on the
PHV are significantly more stringent in the absence of SOC.  We forgo a general discussion
here, but give a simple argument that whenever the applied magnetic
field and all local ordered spin moments lie in a single plane, the
PHV vanishes.  Indeed, under those assumptions, all Zeeman and exchange fields
are invariant under the operation
$\mathcal{T} \left[C_{2\bm{\hat{w}}}\right]_s$, where the second
operation is a $C_2$ rotation {\em in spin space} about an axis
$\bm{\hat{w}}$ normal
to the plane containing the spins and field.  Because spin rotations
do not act on the strain, this operation is indistinguishable from
time-reversal symmetry in the lattice effective action, and thus PHV
is prohibited.  In a real material with weak but nonzero SOC, there
will be an additional smallness of the PHV due to weak SOC.

We now turn to an estimate of the magntitude of the phonon Hall effect
induced by the mechanism in this paper.  The characteristic energy
$v_{\rm ph}\delta_\alpha$ and the maximum of $\kappa_{xy}/T$ can be
related to parameters that may be obtained from experiments/ab-initio
calculations.  The thermal conductivity is most conveniently expressed
as $\kappa_{xy}/T$ per layer and in terms of the ratio of the phonon
to the magnon velocity $\Upsilon=v_{\rm ph}/v_m$. In units of thermal
conductance, we have:
\begin{align}
\frac{d_z
  \kappa^{\Gamma\Gamma'}_{xy}}{T}\sim\frac{\gamma_{\Gamma\Gamma'}d_z}{\rho
  v_{\rm ph}}\frac{k_B^2}{\hbar}= \Upsilon\frac{\lambda_\Gamma
  \lambda_{\Gamma'}}{J \, v_m/d_z}\frac{h}{\rho a_0^2 d_z v_{\rm ph}^2} \frac{k_B^2}{\hbar}\times{O}(1).
\label{eq:kappaval}
\end{align}
Here $d_z$ is the inter-layer distance, and the numerical coefficient
(${O}(1)$) depends on the specific microscopic model.  Because the
magnetoelastic couplings may be considered spatial derivatives of the
(anisotropic) exchange, and the magnon velocity is also set by
exchange, we expect that
$\frac{\lambda_\Gamma \lambda_{\Gamma'}}{J \, v_m/d_z}\lesssim 1$
(smallness due to weak SOC also enters here).  It could be more precisely
evaluated through ab initio calculations.   The remaining factors
behave as
$\frac{h}{\rho a_0^2 d_z v_{\rm ph}^2}\sim h(C_{\rm 3D} a_0^2 d_z)^{-1}$, where
$C_{\rm 3D}$ is the three-dimensional bulk modulus. The latter is
generally an eV energy so that for a magnetic field $B=15$~T,
i.e.\ $h=g_s\mu_B B\sim 30$~K, $h(C_{\rm 3D}a_0^2 d_z)^{-1} \sim
10^{-4}$.   For cuprates, $\Upsilon <1$ due to the large
exchange energy.  Thus, the thermal Hall conductivity due to the PHV
mechanism is smaller by a factor of at least $10^{-4}$ than
measured values for cuprates. 

This leaves open other mechanisms for the phonon Hall effect in the
cuprates.  It is possible that the effects of the PHV are enhanced by
impurity scattering of phonons, not included here (see a recent preprint~\cite{Guo2021} for discussion on related issues).
There might also be alternative mechanisms to generate larger PHV, e.g.\ from charged
impurities.  Finally, phonons may aquire chirality via skew scattering
of phonons from spins.  This mechanism, which is independent of the
PHV, requires a true non-equilibrium transport treatment.

A partial means to distinguish between these possibilities is to
compare directly the acoustic Faraday and phonon Hall effect
measurements to see if they can be consistently related to comparable
PHVs.  Beyond the cuprates, the above discussion can be easily
extended to any antiferromagnet, and may be used to guide a search for
large PHVs.  At a general level, it is clear that large spin-orbit
coupling or non-coplanar magnetic order (which evades the weak SOC
smallness), and large magnetoelastic couplings are beneficial for
enhancing the PHV.  We expect our results will be useful to guide
future experiments and computations in this active area.

\begin{acknowledgments}
We acknowledge L\'eo Mangeolle for a
collaboration on a related project and helpful discussions. M.Y.\ also benefited from collaborations with Natalia Perkins and Rafael Fernandes on a past work on phonon Hall viscosity. L.B. was
supported by the DOE, Office of Science, Basic Energy Sciences under
Award No. DE-FG02-08ER46524.  M.Y.\ is
supported in part by the Gordon and Betty Moore Foundation through
Grant GBMF8690 to UCSB and by the National Science Foundation under
Grant No.\ NSF PHY-1748958. L.S.\ acknowledges funding from the
European Research Council (ERC) under the European Union's Horizon
2020 research and innovation program (Grant agreement No.\ 853116,
``TRANSPORT'') as well as from the ToRe Idex-Lyon breakthrough
program.
\end{acknowledgments}

\bibliography{PhononHallCuprates} 

\clearpage
\onecolumngrid
\appendix
\section*{Supplemental Material}
In the supplemental material (SM), we present additional details of
the modeling and computations.

Sec.~\ref{app:NLSM} reviews the detailed derivation of the correlation function of staggered and ferromagnetic field in Eq.~\eqref{eq:correlator} of the main text based on the non-linear sigma model formulation.
Summaries of the point group symmetry operations and symmetry allowed magnetic operators in the spin-lattice coupling [Eq.~\eqref{eq:Lsl} in main the
text] are given in Sec.~\ref{app:Symmetry}. In Sec.~\ref{app:exp}, we show details to analyze
the acoustic Faraday and phonon thermal Hall effects. 

For completeness, a few additional considerations are presented. We first show that the next order in $1/S$ contribution to $\eta^H$ from two magnon fluctuations are small in magnitude at low temperature, and can thus be ignored (Sec.~\ref{app:twomagnon}). Next, we show the microscopic Hamiltonian allowed by crystal symmetry in Sec.~\ref{app:MicroscopicH}. A comparison between the microscopic Hamiltonian and the low energy effective action analysis is helpful to infer the spin gap and reveal the microscopic origin of the relevant spin-lattice coupling for a material.  Finally, the phonon Hall viscosity with in-plane magnetic field is discussed (Sec.~\ref{app:etaHy}). 
\section{Non-linear sigma model formulation of spin dynamics}
\label{app:NLSM}
In this section, we derive the low energy field theory and
correlations function for the antiferromagnetic state.  Before doing
so, we address briefly the validity of the low energy approach.  Since
$\etaH(\qv)$ arises from virtual rather than on-shell magnetic
excitations, there is no kinetic constraint, and spin fluctuations at
all energies may contribute to it.  However, high energy magnetic
excitations are still polynomially suppressed by their energies, and
indeed we have checked that, within a full spin wave calculation,
contributions from high-energy magnons are suppressed by additional
powers of $k_B T/J$ with respect to the low energy ones.  This
justifies the low energy approach.

We proceed with the standard non-linear sigma model (NLSM)
formulation for collinear antiferromagnets and obtain the two-point
correlation functions for a staggered field ($\ve{n}$) and a
ferromagnetic field ($\bm{m}$) in an external magnetic field ($\bm{h}$) in a two-dimensional (2d) spin system. It gives Eq.~\eqref{eq:correlator} in the main text.

Following Ref.~\cite{SachdevBook}, the coherent state path integral
for a Heisenberg spin with spin value $S$ at site $\ver$ can be
obtained in the basis of the (unit) vector field $\ve{e}_\ver$, which
is defined through
$\vhat{S}_\ver\ket{\ve{e}_\ver}=S \ve{e}_\ver \ket{\ve{e}_\ver}$ and
$|\ve{e}_\ver|^2=1$.  For a collinear antiferromagnet, the ground
state has spins oriented in opposite directions on the two sublattices
(defined as $A$, $B$ sublattice hereafter). The low energy spin
dynamics can be described by a set of continuous fields, which include
the staggered $\ve{n}\sim \ve{e}_{A}-\ve{e}_B$ and uniform
$\bm{m}\sim \ve{e}_A+\ve{e}_B$ magnetization fields, where the
subscript $A,\, B$ label the A, B sublattice. To be accurate, we use
\begin{align}
\ve{e}_\ver=(-1)^\ver \ve{n}_{\ver} \sqrt{1-(\bm{m}_{\ver})^2}+\bm{m}_{\ver},
    \label{appeq:NLSMN}
\end{align}
where $\bm{m}$ is the uniform magnetization per site in units of the
saturation magnetization ($=S$ semiclassically), and $(-1)^\ver$ is a sign equal to $+1$ on the A sublattice and
$-1$ on the B sublattice.    The effective spin action (for the isotropic Heisenberg model) is
\begin{align}
    \mac{Z}_s &= \int \Diff\ve{n}\Diff \bm{m} \delta(\ve{n}^2-1)\delta(\ve{n}\cdot\bm{m})\exp{(-\mac{S}_s)}\non\\
    \mac{S}_s & =\frac{1}{2}\int\diff x \diff y \diff
                \tau\{\frac{v_m}{ga_0}[(\ve{\nabla}_x\ve{n})^2+(\ve{\nabla}_y\ve{n})^2]+\frac{v_m
                g S^2}{a^3_0}\, \bm{m}^2-\frac{2 i S }{a_0^2} \bm{m}\cdot \left(\ve{n}\times \frac{\partial \ve{n}}{\partial \tau}-i \ve{h}\right)\},
    \label{appeq:spinaction}
\end{align}
In the second equation, we introduced the spin wave (magnon) velocity $v_m\sim Ja_0 S$, with $a_0$ the
lattice constant, and the coefficient of $\ve{m}^2$ defines
the coupling $g$ with $g\sim S^{-1}$.   We ignore the spin wave velocity along
$\vhat{z}$ because of the much weaker interlayer spin exchange. In
the standard procedure the Gaussian field $\bm{m}$ is then integrated out. As $\bm{m}$ does not have its own dynamics, the action can be obtained by replacing $\bm{m}$ with its saddle-point solution, 
\begin{align}
     \bm{m} = \chi\left(i\ve{n}\times \frac{\partial \ve{n}}{\partial \tau}+ \ve{h}-\ve{n}(\ve{n}\cdot\ve{h})\right), 
    \label{appeq:Lsaddle}
\end{align}
which defines the susceptibility $\chi = a_0/v_mgS$.

Below, we consider an antiferromagnet with the N\'eel  vector along
the $x$-axis, i.e.\ $\lim_{h\rightarrow 0} \langle \bm{n}\rangle =
n_0\bm{\hat{x}}$, and the magnetic field perpendicular to the N\'eel
vector. As the spin-lattice coupling is most transparently expressed
in terms of $\ve{n}$, $\bm{m}$ fields to reveal the symmetries, our
goal below is to obtain the correlators of $\ve{n}$ and $\bm{m}$. The
staggered field can be parameterized by
$\ve{n}=\{n_0,n_y,n_z\}=(n_0,\ve{n}_\pp)$, where $n_0=\sqrt{
  1-\ve{n}_\pp^2}$ is the order parameter, and $n_{y,z}$ are
transverse fluctuations (spin waves). Note from
Eq.~\eqref{appeq:Lsaddle} that $\bm{m}$ can be decomposed into
components of zeroth order in $\ve{h}$ and first order in
$\ve{h}$. The first-order term includes the static part, i.e.\ the
field induced uniform magnetization $\chi \ve{h}$ as well as $-\chi \ve{n}(\ve{n}\cdot\ve{h})$ to satisfy
the constraint $\bm{n}\cdot \bm{m}=0$ at first order in $\ve{h}$. We
then define $\bf{m}$ through $\bm{m}=\chi \ve{h}+\bf{m}$, and within
linear spin wave theory, ${\rm m}_x=-\chi n_x(\ve{n}\cdot\ve{h}) \rightarrow -\chi n_0 (n_y h_y + n_z h_z)$. Plugging $\chi \ve{h}$ and ${\rm m}_x$ to $\mac{S}_s$ into Eq.~\eqref{appeq:spinaction}, we obtain the Lagrangian density in terms of $\ve{n}_\pp$ and $\bf{m}_\pp$. 

\paragraph{With an external field $\ve{h}=h \bm{\hat{z}}$.}   
    \begin{align}
    \mac{S}_s = \frac{1}{2} \sum_{\kv,n} 
    \begin{pmatrix}
    n_y & {\rm m}_y & n_z & {\rm m}_z
    \end{pmatrix}_{\kv,n}
    \begin{pmatrix}
    \frac{1}{gv_m a_0} (v_m^2\bm{k}^2+\Delta_y^2) & & & -\frac{S\omega_n}{a_0^2} \\
    & \frac{v_mgS^2}{a_0^3} & -\frac{S\omega_n}{a_0^2} & \\
    & \frac{S\omega_n}{a_0^2} & \frac{1}{gv_m a_0} (v_m^2\bm{k}^2+h^2+\Delta_z^2) & \\
    \frac{S\omega_n}{a_0^2} & & & \frac{v_mg S^2}{a_0^3}\\
    \end{pmatrix}
     \begin{pmatrix}
    n_y\\
    {\rm m}_y\\
    n_z\\
    {\rm m}_z\\
    \end{pmatrix}_{-\kv,-n}
    \label{appeq:action}
    \end{align}
    Here, the Fourier transform follows the convention
    $f_{\kv}=\frac{1}{L} \int \diff x \diff y \,e^{-i\kv\cdot \ve{x}}
    f(\ve{x})$, where $L$ is the linear size of the sample
    ($\int \!dx dy\, 1 = L^2$).  Inverting the matrix, we obtain the
    correlator:
     \begin{align}
    \langle T_\tau 
    \begin{pmatrix}
    n_\alpha \\
    {\rm m}_{\bar{\alpha}}
    \end{pmatrix}_{\bm{k}}
    \begin{pmatrix}
    n_\alpha & {\rm m}_{\bar{\alpha}}
    \end{pmatrix}_{-\bm{k}}\rangle_{\omega_n}
    =\mac{D}_\alpha(\bm{k},\omega_n)
    \begin{pmatrix}
    v_m g a_0 & \frac{\omega_na_0^2}{S}\,\epsilon_{x\alpha\bar{\alpha}}\\
    -\frac{\omega_na_0^2}{S}\,\epsilon_{x\alpha\bar{\alpha}} &
    \frac{\upomega^2_{\alpha,\bm{k}}a_0^3}{v_m g S^2}
    \end{pmatrix}.
    \label{appeq:correlator}
\end{align}
Here $a_0$ is the lattice constant,
$\mac{D}_\alpha^{-1}(\bm{k},\omega_n)=\omega_n^2+\upomega_{\alpha,\bm{k}}^2$,
$\upomega_{\alpha,\bm{k}}$ is the dispersion for the magnon branch $\alpha\in
\{y,z\}$, such that
$\upomega_{\alpha,\bm{k}}=v_m\sqrt{k_x^2+k_y^2+\delta_\alpha^2}$, the spin
gap is determined by the XYZ anisotropy ($\sim \Delta_\alpha$) and
external field strength $h$ by $\delta_y=\Delta_y/v_m$,
$\delta_z=\sqrt{\Delta_z^2+h^2}/v_m$. 

    \paragraph{With an external field $\ve{h}=h \bm{\hat{y}}$.}
    Because here also
    $\ve{h}\perp \langle \ve{n}\rangle$, the situation is similar to
    that when
    $\ve{h}=h\bm{\hat{z}}$. The effective spin action simply changes
    according to $\Delta_y^2\rightarrow (\Delta_y^2+h^2)$, $\Delta_z^2+h^2\rightarrow \Delta_z^2$ in Eq.~\eqref{appeq:action}. Consequently, $\delta_y=\sqrt{\Delta_y^2+h^2}/v_m$ and $\delta_z=\Delta_z/v_m$.

\subsection{Normalization of operators, Fourier conventions, etc.}
\label{sec:norm-oper-four}

Here we discuss the conventions used to obtain the Hall viscosity form
given in the main text, Eq.~\eqref{eq:eta2}.  We begin with a
consideration of units.  Hall viscosity, Eq.~\eqref{eq:PHV}, is
defined as a coefficient in a three-dimensional elastic theory.  We
employ Fourier conventions for the strain which are appropriate to a
three-dimensional system, so that
\begin{equation}
  \label{eq:6}
  \mathcal{E}_{\Gamma,\bm{q}} =\frac{1}{\sqrt{V}} \int\! d^3\bm{x}\,
  \mathcal{E}_\Gamma(\bm{x}) e^{-i\bm{q}\cdot\bm{x}}, \qquad
  \mathcal{E}_\Gamma(\bm{x}) = \frac{1}{\sqrt{V}} \sum_{\bm{q}}
  e^{i\bm{q}\cdot\bm{x}} \mathcal{E}_{\Gamma,\bm{q}},
\end{equation}
where $V$ is the volume of the system.  With this convention, since
the real space strain $\mathcal{E}_\Gamma(\bm{x})$ is dimensionless,
$\mathcal{E}_{\Gamma,\bm{q}}$ has dimensions of $1/\sqrt{V}$.  Then
Eq.~\eqref{eq:PHV} implies that, because the action is
dimensionless, $\eta^H$ has units of inverse volume.

Next consider the spin-lattice coupling in Eq.~\eqref{eq:Lsl}.  The
Lagrange density has units of energy density, so that the combination
$\lambda_\Gamma \mathcal{O}_\Gamma$ must have units of energy density.
We assigned energy units to $\lambda$, which requires
$\mathcal{O}_\Gamma$ to scale as a number density.  How this is
precisely realized depends upon our treatment of the third dimension.
We will proceed here with the treatment as a discrete layered system,
so that the corresponding action is
$\mathcal{S}_{sl}=\sum_z \int\! dx dy\, \mathcal{L}_{sl}$.  Then
$\mathcal{O}_\Gamma$ has units of inverse length squared.  Let us see
how these factors appear in the derivation of Eq.~\eqref{eq:eta2}.

All the contributions to the spin-lattice coupling arise
microscopically from expressions (see Sec.~\ref{app:MicroscopicH}) of
the form $\sum_\ver \lambda_\Gamma \mathcal{E}_\Gamma (SS)_{\Gamma,\ver}$,
where $(SS)_{\Gamma,\ver}$ represents a sum of spin bilinears in the vicinity of
site $\ver$.  We convert this to continuum fields using
Eq.~\eqref{appeq:NLSMN} and $\sum_\ver \rightarrow a_0^{-2}\sum_z\int\!
dx dy\,$, which gives
\begin{equation}
  \label{eq:7}
  \mathcal{O}_{B_1} = \frac{S^2 \chi h {\rm m}_z}{a_0^2}, \qquad
  \mathcal{O}_{B_2} = \frac{S^2 n_0 n_y}{a_0^2}, \qquad
  \mathcal{O}_{E_x} = \frac{S^2 n_0 n_z}{a_0^2}, \qquad
  \mathcal{O}_{E_y} = \frac{S^2 \chi h {\rm m}_y}{a_0^2}.
\end{equation}
The factors of $S^2,\chi, 1/a_0^2,n_0$ are subsumed in the $\sim$ in the
discussion of the main text.  
Now consider evaluating Eq.~\eqref{eq:1}.  We must take care due to
the combination of the three-dimensional Fourier transform convention
for elasticity with our two-dimensional magnetism theory.   Consider
the $B_1-B_2$ contribution.  What actually arises in the effective
action is
\begin{equation}
  \label{eq:8}
  \mathcal{S}_{sl}^{B_1 B_2} = \sum_{z,z'} \int\! dx dy d\tau\,\int\! dx' dy' d\tau'\,  \lambda_{B_1}
  \lambda_{B_2} \mathcal{E}_{B_1}(x,y,z,\tau) \mathcal{E}_{B_2}(x',y',z',\tau')
  \langle \mathcal{O}_{B_1}(x,y,z,\tau) \mathcal{O}_{B_2}(x',y',z',\tau')\rangle.
\end{equation}
Since we assume no spin correlations between layers, the summand is
non-zero only for $z'=z$.  Inserting the three-dimensional Fourier
expression of Eq.~\eqref{eq:6} for the strains gives
\begin{equation}
  \label{eq:9}
  \mathcal{S}_{sl}^{B_1 B_2} = \frac{1}{V}\sum_{z} \sum_{\bm{q},\bm{q}'} \int\! dx dy d\tau\,\int\! dx' dy' d\tau'\,  \lambda_{B_1}
  \lambda_{B_2} \mathcal{E}_{B_1,\bm{q}}(\tau)
  \mathcal{E}_{B_2,\bm{q}'}(\tau')
  e^{i(\bm{q}_\perp\cdot\bm{x}_\perp+\bm{q}'_\perp\cdot\bm{x}'_\perp)} e^{i(q_z+q'_z)z}
  \langle \mathcal{O}_{B_1}(x,y,\tau) \mathcal{O}_{B_2}(x',y',\tau')\rangle.
\end{equation}
The sum over $z$ gives $N_z \delta_{q_z+q'_z,0}$, where $N_z=L_z/d_z$ is
the number of layers.  Combining this with the $1/V$ prefactor gives
$\frac{L_z}{d_z}\frac{1}{V} = \frac{1}{d_z L^2}$.  We can now use the
$1/L^2$ to form the prefactors of the {\em two-dimensional} Fourier
transform for each of the two $\mathcal{O}$ operators.  Hence
\begin{equation}
  \label{eq:10}
   \mathcal{S}_{sl}^{B_1 B_2} = \sum_{\bm{q}_\perp,\bm{q}'_\perp}\sum_{q_z} \frac{\lambda_{B_1}
  \lambda_{B_2}}{d_z}  \int \! d\tau d\tau'\, \mathcal{E}_{B_1,\bm{q}}(\tau)
  \mathcal{E}_{B_2,\bm{q}'}(\tau') \langle
  \mathcal{O}_{B_1,-\bm{q}_\perp}(\tau) \mathcal{O}_{B_2,-\bm{q}'_\perp}(\tau')\rangle
\end{equation}
Due to momentum conservation, $\bm{q}'_\perp=-\bm{q}_\perp$ is the only
non-zero correlator, and we obtain  using Eq.~\eqref{eq:7} finally
\begin{equation}
  \label{eq:11}
  \eta^H_{B_1,B_2}(\bm{q}) = \frac{\lambda_{B_1}\lambda_{B_2}}{d_z}
  \frac{S^2\chi h}{a_0^2} \frac{S^2 n_0}{a_0^2} \times \left[ -i
    \partial_\omega \langle {\rm
    m}_{z,-\bm{q}_\perp}
  n_{y,\bm{q}_\perp}\rangle_{\omega_n\rightarrow -i\omega+0^+} \right].
\end{equation}
Inserting Eq.~\eqref{eq:correlator} for the ${\rm m}_z-n_y$
correlation function inside the square bracket, one obtains the result
in Eq.~\eqref{eq:eta2} for $\Gamma=B_1$, $\Gamma'=B_2$.

We took some
pains to present this in great detail for clarity, but the result can also be
understood schematically on dimensional grounds: the three-dimensional
Fourier tranform differs from the two-dimensional one by a factor of
the inverse of the square root of a length in the $z$ direction.
Converting the 2d to 3d Fourier conventions for the two $\mathcal{O}$
operators appearing at second order in the spin-lattice coupling, one
obtains an overall factor of $1/d_z$.  This is the factor in the first
term in Eq.~\eqref{eq:11}.  The remaining factors were explained
previously as arising from the conversion from the lattice to the 2d
continuum theory.

We note furthermore that it would have been possible to formulate the
magnetic correlations in three dimensions as well, which is in a sense
more general, and would also avoid some of this confusion.  We opted
for the present formulation in order to emphasize, as discussed in the
main text, that the results apply to any three-dimensional structure
composed of such 2d layers, and that no three-dimensional magnetic
correlations are required to induce the desired PHV terms, even those
which involve the inherently three-dimensional $E_x,E_y$ strains.

\section{Group theory analysis}
\label{app:Symmetry}
Here, the relevant symmetry groups and symmetry operations on the strain and spin fields are listed for reference. We consider the example discussed in the main text, i.e.\ a crystal with layer group symmetry $P4/mmm$ (number 61 of layer group).  The symmetry allowed magnetic operators that appear in the spin-lattice coupling [Eq.~\eqref{eq:Lsl} in the main text] are also presented in Tabs.~\ref{tab:myops} and~\ref{tab:shifted}.

\subsection{List of point group symmetry operations}
\begin{figure}[h]
    \centering
    \includegraphics[width=0.35\columnwidth]{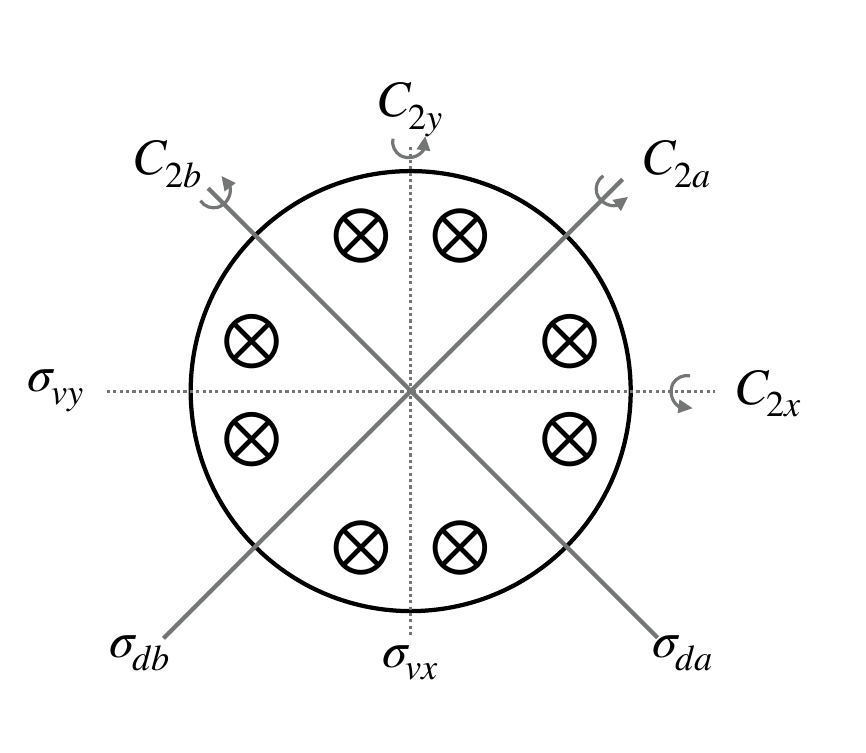}
    \caption{Point group symmetry transformations for ${\rm D}_{4h}$. The label $\otimes$ denotes horizontal mirror symmetry $\sigma_h$ that relates $\times$ and $\circ$.  }
    \label{fig:sym}
\end{figure}

The generators of the point group symmetry for the underlying crystal,
${\rm D}_{4h}$, are ${\rm D}_{4h}=\langle C_{4z}, \sigma_h, \sigma_v\rangle$. They
give 16 point group elements: ${\rm D}_{4h}=\{{\rm Id}, 2 C_{4z} , C_{2z} ,
2C_{2}', 2 C_{2}'', i , \sigma_h , 2\sigma_v , 2\sigma_d , 2S_4
\}$~\cite{Koster1963}. Here ${\rm Id}$ denotes the identity. $2C_2'$ denotes the two $\pi$ rotation transformations,
around the $\bm{\hat x}$ or $\bm{\hat y}$ axes, respectively. $2C_2''$ denotes the two $\pi$ rotations
around the diagonal axes $\bm{\hat a}=\frac{1}{\sqrt{2}}(\bm{\hat
  x}+\bm{\hat y})$ and $\bm{\hat b}=\frac{1}{\sqrt{2}}(-\bm{\hat x}+\bm{\hat y})$, respectively. $\sigma_v$ and
$\sigma_d$ are mirror transformations with a mirror plane
perpendicular to the $xy$ plane, whose normal direction is along
$\bm{\hat x},\bm{\hat y}$ and diagonal axis $\bm{\hat a},\bm{\hat b}$,
respectively. $\sigma_h$ is a horizontal mirror reflection. The convention is shown in
Fig.~\ref{fig:sym}, such that the inversion transformation $i$ can be
obtained from $i=C_{2\alpha}\sigma_\alpha$, where
$C_{2\alpha}=C_{2x,2y,2z,2a,2b}$,
$\sigma_\alpha=\sigma_{vx,vy,h,da,db}$. Here, $\alpha=x,y,z,a,b$ denotes the axis of two-fold rotation for $C_{2\alpha}$, and the axis is also the normal vector of mirror plane for the respective $\sigma_{\alpha}$, with $\alpha=vx,vy,h,da,db$.  

An external field $\ve{h}=h\bm{\hat{z}}$ breaks the point group
${\rm D}_{4h}$ down to ${\rm C}_{4h}=\langle C_{4z}, \sigma_h\rangle=\{E, 2 C_{4z}
, C_{2z} , i , \sigma_h , 2S_4\}$ through the Zeeman term
$H_Z=\ve{h}\cdot \sum_\ver \ve{S}_\ver$. In terms of the antiunitary
symmetries, $\trl$ is broken, while the Zeeman term is invariant under
the antiunitary symmetries $2\trl{C_2'}\oplus 2\trl{C_2''} \oplus 2\trl{\sigma_v}  \oplus 2 \trl{\sigma_d}$. This gives the black-white magnetic point group $4/mm'm'=\left({\rm C}_{4h}+ \trl\times ({\rm D}_{4h}-{\rm C}_{4h})\right)$ as listed in Tab.~I.

An external field $\ve{h}=h\bm{\hat{y}}$ 
breaks ${\rm D}_{4h}$ down to ${\rm C}^{\prime}_{2h}=\langle C_{2y}, i\rangle=\{E,
C_{2y} , i , \sigma_{vy}\}$, whose abstract group structure is the
same as that of ${\rm C}_{2h}$. It also preserves the antiunitary symmetries $\trl{C_{2z}}\oplus \trl{\sigma_h} \oplus \trl{\sigma_{vx}} \oplus \trl{C_{2x}}$, which can be generated by e.g.\ $\trl{C_{2z}}$ and ${\rm C}^{\prime}_{2h}$. 

\subsection{Symmetry operations}
The point group symmetry operations acting on the strain field $\mathcal{E}_\Gamma$, the spin vector $\ve{S}$, the lattice coordinate $\{x,y,z\}$, the continuous spin fields $\ve{n}, \bm{m}$ are listed in Table.~\ref{tab:SymTransform}. For brevity, only the point group symmetry generators listed in Tab.~\ref{tab:MagneticSG} and Tab.~\ref{tab:MagneticSGy} are shown.
\begin{table}[th]
\begin{center}
\begin{tabular}{|c||c|c|c||c||c|c|c|}
\hline
 & $C_{4z}$ &  $\sigma_{vx}$ & $\sigma_{h}$ & $i$ & $C_{2z}$ & $C_{2y}$ & $C_{2x}$ \\
\hline
%
$\mathcal{E}_{xx}+\mathcal{E}_{yy}$ ($\mathcal{E}_{A_1}$) & & & & & & & \\ 
$\mathcal{E}_{xx}-\mathcal{E}_{yy}$ ($\mathcal{E}_{B_1}$) & $-\mathcal{E}_{xx}+\mathcal{E}_{yy}$ &  & &  & & & \\ 
$\mathcal{E}_{xy}$ ($\mathcal{E}_{B_2}$) &  $-\mathcal{E}_{xy}$ & $-\mathcal{E}_{xy}$ & &  & & $-\mathcal{E}_{xy}$ &$-\mathcal{E}_{xy}$\\ 
$\{\mathcal{E}_{xz},\,\mathcal{E}_{yz}\}$ ($\mathcal{E}_{E}$) &  $\{-\mathcal{E}_{yz},\,\mathcal{E}_{xz}\}$ & $\{-\mathcal{E}_{xz},\,\mathcal{E}_{yz}\}$ & $\{-\mathcal{E}_{xz},\,-\mathcal{E}_{yz}\}$ &  & $\{-\mathcal{E}_{xz},\,-\mathcal{E}_{yz}\}$ & $\{\mathcal{E}_{xz},\,-\mathcal{E}_{yz}\}$ & $\{-\mathcal{E}_{xz},\,\mathcal{E}_{yz}\}$ \\ 
\hhline{|--------|}
$\{S_x,S_y\}$ &  $\{- S_y,S_x\}$ & $\{S_x,-S_y\}$ & $\{-S_x,-S_y\}$ & & $\{-S_x,-S_y\}$ & $\{-S_x,S_y\}$ & $\{S_x,-S_y\}$\\
$S_z$ &  &  $-S_z$ &  &  &    & $-S_z$ & $-S_z$  \\
\hhline{|--------|}
$\{x,y\}$ &  $\{-y,x\}$ & $\{-x,y\}$ &  & $\{-x,-y\}$ & $\{-x,-y\}$ & $\{-x,y\}$ & $\{x,-y\}$ \\
$z$ & &  & $-z$ & $-z$   &  & $-z$ & $-z$  \\
\hhline{|--------|}
$n_x/m_x$ & $-n_y/-m_y$ &  & $-n_x/-m_x$ & & $-n_x/-m_x$ & $-n_x/-m_x$ &  \\
$n_y/m_y$ & $n_x/m_x$ & $-n_y/-m_y$ & $-n_y/-m_y$ & & $-n_y/-m_y$ &  & $-n_y/-m_y$ \\
$n_z/m_z$ &  & $-n_z/-m_z$ &  & & & $-n_z/-m_z$ & $-n_z/-m_z$ \\
\hline
\end{tabular}
\caption{\label{tab:SymTransform} Symmetry transformation of the elastic strain tensor, spin vector, lattice coordinates, and continuous spin fields. The first row lists the important point group transformations to find the magnetic space group generators discussed above. The blank space in the table denotes the variable is invariant under the corresponding transformation. Note that from the relations
$i C_{2\alpha} = \sigma_\alpha$ and
$C_{2\alpha}C_{2\beta}=C_{2\gamma}, C_{2\alpha}\sigma_\beta=\sigma_\gamma$, where $C_{2\alpha}=C_{2x,2y,2z,2a,2b}$, $\sigma_\alpha=\sigma_{vx,vy,h,da,db}$ and $\alpha,\beta,\gamma$ are mutually orthogonal basis, other symmetry transformations not listed can be generated.
}
\end{center}
\end{table}

\subsection{Symmetry allowed spin-lattice coupling}
Based on Table~\ref{tab:SymTransform}, it is straightforward to
classify polynomials of $n_\mu, m_\mu$ that couples to the strain field by irreps.  This is given in
Table~\ref{tab:myops}.  Note that we restricted our list to terms without
spatial derivatives, as these suffer additional suppression by
temperature factors well below the Debye
temperature. Furthermore, we kept terms only to linear order in $m_x$,
because in the spin wave expansion, $m_x$, being longitudinal, is
already quadratic in the low energy transverse fields at zeroth order in $h$.  

\begin{table}[h!]
  \def\arraystretch{1.3}
  \begin{center}
    \begin{tabular}{c|| c c c c}
    & $n$ & $n n $ & $m m$ & $n m m$\\
    \hline
    $\mac{O}_{B_1}$  & & $n_y n_y, n_z n_z$ & $m_y m_y, m_z m_z$ & $n_y m_x m_y$,$n_z m_x m_z$ \\
    $\mac{O}_{B_2}$  & $n_y$ &   & $m_x m_y$  & $n_y m_{y/z} m_{y/z},n_z m_y m_z$ \\
    $\mac{O}_{E_x}$  & $n_z$ &   & $m_x m_z$  & $n_z m_{y/z} m_{y/z}, n_y m_y m_z$ \\
    $\mac{O}_{E_y}$  & & $n_y n_z$ & $m_y m_z$ & $n_y m_x m_z$,$n_z m_x m_y$ \\
    \end{tabular}
    \caption{  \label{tab:myops} Operators arranged by irrep for the high symmetry antiferromagnet such that Eq.~\eqref{eq:Lsl} in the main text is invariant under $\mathsf{G}(\vhat{x},\ve{0})$. Other factors, e.g. $S^2,\chi, 1/a_0^2,n_0$, have been omitted in the table.}
  \end{center}
\end{table}

As described in the text, the operators can be expressed in terms of
$\bm{n}_\pp$ and $\bf{m}_\pp$ using the NLSM constraints, after taking into account an external Zeeman
field $\bm{h}$.  This leads to the forms in Table~\ref{tab:shifted} by replacing $m_z=\chi h+{\rm m}_z$, $m_x={\rm m}_x=-\chi n_z h$.

\begin{table}[h!]
\def\arraystretch{1.3}
    \begin{tabular}{c|| c c c c}
    & $n$ & $ n n $ & $ m m $ & $ n m m$\\
    \hline
    $\mac{O}_{B_1}$  & & $n_y n_y, n_z n_z$ & $ h\,{\rm m}_z $ & \\
    $\mac{O}_{B_2}$  & $n_y$ &   & $-h\, n_z {\rm m}_y$  & $ h\, n_y {\rm m}_z , h\, n_z {\rm m}_y $ \\
    $\mac{O}_{E_x}$  & $n_z$ &   & $-h\, n_z {\rm m}_z$  & $ h\, n_z {\rm m}_z, h\, n_y {\rm m}_y $ \\
    $\mac{O}_{E_y}$  & & $n_y n_z$ & $h\,{\rm m}_y $ & \\
    \end{tabular}
    \caption{Magnetic operators up to quadratic order in the transverse spin wave fluctuations in the high symmetry AFM in the
      presence of a small field along the $z$-axis. Other factors, e.g. $S^2,\chi, 1/a_0^2,n_0$, have been omitted in the table. See Eq.~\eqref{eq:7} for the complete expression of $\mac{O}$ at linear order in $n_\mu, {\rm m}_\mu$. \label{tab:shifted}}
\end{table}


\section{Experimental Implications}
\label{app:exp}
Our starting point is the effective phonon Lagrangian. In Fourier space, it reads $\mac{L}_{ph}=\sum_\qv \mac{L}^{(0)}_{ph}(\qv)+\mac{L}_{\rm PHV}(\qv)$, where
\begin{align}
    \mac{L}_{ph}(\qv)= &\frac{1}{2}\rho \dot{\ve{u}}^T_{-\qv}\dot{\ve{u}}_{\qv}-\frac{1}{2}\ve{u}^T_{-\qv}\mathsf{M}_\qv\ve{u}_{\qv}+\dot{\ve{u}}^T_{-\qv}\mathsf{A}_\qv\ve{u}_{\qv}.
    \label{appeq:Lph}
\end{align}
Here, $\rho$ is the lattice mass density, the sans serif font denotes a matrix in Euclidean space,
the first and second terms are the harmonic acoustic phonon Lagrangian $\mac{L}^{(0)}_{ph}$, and the third term comes from the phonon Hall viscosity and
$\mathsf{A}_\qv=-\mathsf{A}_\qv^T$. For simplicity, we will ignore the
anisotropy in $\mac{L}^{(0)}_{ph}$ in the evaluation of the phonon Berry curvature. $\mathsf{M}_\qv$ for a 3D isotropic elastic medium is
\begin{align}
\mathsf{M}_{\bm{q}}=\begin{pmatrix}
c_1 q_x^2+c_2 q^2 & c_1 q_x q_y & c_1 q_x q_z \\
c_1 q_x q_y & c_1 q_y^2 + c_2 q^2 & c_1 q_y q_z \\
c_1 q_x q_z & c_1 q_y q_z & c_1 q_z^2 + c_2 q^2 \\
\end{pmatrix}.
\label{appeq:Mq}
\end{align}
Here $c_{1,2}$ are the elastic modulus tensor
coefficients. The eigenmodes
include two degenerate transverse acoustic waves with sound wave
velocity $v_T=\sqrt{c_2/\rho}$, and one longitudinal wave with
$v_L=\sqrt{(c_1+c_2)/\rho}$. However, the anisotropy of a crystal with
lower symmetry breaks the degeneracy and mixes transverse and
longitudinal waves at a generic momentum.

 The antisymmetric
$\mathsf{A}$ matrices for $\eta^H_{B_1 B_2}$, $\eta^H_{E_x E_y}$ (relevant when the field is along z-axis) are
\begin{align}
\mathsf{A}_\qv^{B_1 B_2}=\eta^H_{B_1 B_2}(\qv)\begin{pmatrix}
0 & (q_x^2+q_y^2) & 0\\
-(q_x^2+q_y^2) & 0 & 0\\
0 & 0 & 0\\
\end{pmatrix},\qquad
\mathsf{A}_\qv^{E_x E_y}=\eta^H_{E_x E_y}(\qv)\begin{pmatrix}
0 & q_z^2 & q_y q_z\\
-q_z^2 & 0 & -q_x q_z\\
-q_y q_z & q_x q_z & 0\\
\end{pmatrix},
  \label{eq:SGdef}
\end{align}
where $\eta^H_{\Gamma \Gamma'}(\qv)$ have been obtained in the main
text. We reproduce the result here: $\eta^H_{B_1 B_2}(\qv)=\frac{\gamma_{B_1 B_2}}{(q_x^2+q_y^2)+\delta_y^2}$, $\eta^H_{E_x E_y}(\qv)=\frac{\gamma_{E_x E_y}}{(q_x^2+q_y^2)+\delta_z^2}$, where $\delta_y=\Delta_y/v_m$, $\delta_z=\sqrt{\Delta_z^2+h^2}/v_m$, and $\gamma_{\Gamma\Gamma'}= \frac{hS^2\lambda_\Gamma \lambda_{\Gamma'}
  }{v^3_m g  d_z a_0}$. 

\subsection{Acoustic Faraday effect}
\label{app:AFE}

The acoustic Faraday effect can be observed in an anisotropic medium
only when the acoustic wave is propagating along a high symmetry
direction, such that the transverse waves remain degenerate and do not mix with the longitudinal one. In our case, for a tetragonal lattice crystal with ${\rm D}_{4h}$ point group symmetry, there are indeed two degenerate transverse modes and one longitudinal mode at $\ve{q}=q_z \bm{\hat{z}}$, due to the $C_{4z}$ symmetry in the little group of the high symmetry line. Upon applying a magnetic field along $\bm{\hat{z}}$, the magnetoelastic-coupling-induced phonon Hall viscosity in the $\eta^H_{E_x E_y}$ channel lifts the degeneracy, and the left/right circularly polarized components are eigenmodes and non-degenerate. 

To be specific, the Lagrangian of a wave with frequency $\omega$ propagating along $\vhat{z}$ with $\ve{q}=q\bm{\hat{z}}$ can
be expressed as:
\begin{align}
\mac{L}_{ph}(q\bm{\hat{z}})=\frac{1}{2}\ve{u}^T_{-\qv}
\begin{pmatrix}
\rho \omega^2-c_2 q^2 & i2\eta^H_{E_x E_y}q^2\omega &  \\
-i2\eta^H_{E_x E_y}q^2\omega & \rho \omega^2-c_2 q^2 &  \\
 &  & \rho \omega^2-(c_1+c_2) q^2 \\
\end{pmatrix}\ve{u}_{\qv}.
\label{appeq:Lafe}
\end{align}
While Eq.~\eqref{appeq:Lafe} is obtained from Eq.~\eqref{appeq:Lph},
which is only valid for an isotropic medium, it also describes the
acoustic phonon along $\bm{\hat{z}}$ for a tetragonal group. The
transverse modes can be diagonalized as left
($\bm{\hat{e}}_+=\frac{1}{\sqrt{2}}\{1,i,0\}^T$) and right
($\bm{\hat{e}}_-=\frac{1}{\sqrt{2}}\{1,-i,0\}^T$)  circularly-polarized
waves, with $\rho \omega^2-c_2 q_{\pm}^2\mp2\eta^H_{E_x
  E_y}q_{\pm}^2\omega=0$. Consequently, the transverse
linearly-polarized wave along $\bm{\hat{z}}$ at frequency $\omega_{\rm
  ph}$ undergoes a Faraday rotation. The Faraday rotation angle per
unit length, which is given by the difference of the left and right
wave numbers, $q_+$ and $q_-$ respectively, is
\begin{align}
\frac{\Phi}{L}=\frac{1}{2}(q_+-q_-)=\frac{\omega_{\rm ph}^2\eta^H_{E_x E_y}(q\bm{\hat{z}})}{v_T^3\rho}+{O}(\omega_{\rm ph}^3),
\end{align}
where $v_T=\sqrt{c_2/\rho}$ is the transverse sound wave velocity along $\bm{\hat{z}}$. Note that the magnon spectrum is dispersionless along $q\bm{\hat{z}}$ due to the weak interlayer spin interactions, so $\eta^H_{E_x E_y}(q\bm{\hat{z}})=\frac{\gamma_{E_x E_y}}{\delta_z^2}$ from Eq.~\eqref{eq:eta2}. This gives Eq.~\eqref{eq:AFE} in the main text.

\subsection{Phonon thermal Hall conductivity}
\label{app:PHE}
In this section, we show more details of the calculation of $\kappa^H(T)$. 
To compute the Berry curvature of the phonon bands, we first construct the
effective phonon Hamiltonian $H_{ph}^{\text{eff}}$ from the Legendre transformation of the Lagrangian $\mac{L}_{ph}$. The equation of motion for the pair of canonical variables $\psi_\qv=\{\ve{u}_\qv,\ve{p}_\qv\}^T$ can be obtained from $\dot{\psi}_\qv=-i[\psi_\qv,H_{ph}^{\text{eff}}]$ ($\hbar=1$). This gives the band Hamiltonian $\mac{H}_\qv$ in the basis of $\psi_\qv$. The phonon band energies/eigenstates can then be computed as eigenvalues/eigenvectors of $\mac{H}_\qv$. 

To be specific, we find
\begin{align}
    H_{ph}^{\text{eff}}&=\sum_\qv\frac{\left(\ve{p}_{-\qv}-\ve{A}_{-\qv} \right)\left(\ve{p}_{\qv}-\ve{A}_{\qv} \right)}{2\rho}+\frac{1}{2}\ve{u}^T_{-\qv}\mathsf{M}_\qv\ve{u}_{\qv}
\end{align}
where the canonical momentum is defined as $\ve{p}_{\qv}=\frac{\partial \mac{L}_{ph}}{\partial \dot{\ve{u}}_{-\qv}^i}=\rho \dot{\ve{u}}_{\qv}+\mathsf{A}_\qv \ve{u}_\qv=\rho \dot{\ve{u}}_{\qv}+\ve{A}_\qv$. We then obtain the equation of motion in matrix form as $\dot{\psi}_\qv=-i\mac{H}_\qv\psi_\qv$, where
\begin{align}
    \mac{H}_\qv=\frac{i}{\rho}
    \begin{pmatrix}
     -\mathsf{A}_\qv & \mathbb{I}\\
     -\rho \mathsf{M}_\qv & - \mathsf{A}_\qv
    \end{pmatrix}.
\end{align}
Defining $\xi^{L(R)}_{\qv,\sigma}$ as the left/right eigenvector of
$\mac{H}_\qv$ for the phonon branch $\sigma$ with eigenvalue/spectrum $\upomega_{\qv,\sigma}$, the Berry curvature is
\begin{align}
    \varOmega_{\qv,\sigma}^i=-\im \left[\epsilon_{ijk}\partial_{q_j} \xi^L_{\qv,\sigma} \partial_{q_k} \xi^R_{\qv,\sigma} \right].
    \label{appeq:Berry}
\end{align}
As $\mac{H}_\qv$ is not Hermitian, $\xi^{L(R)}_{\qv,\sigma}$ are not Hermitian conjugate with each other, and they are normalized with $\xi^L_{\qv,\sigma'}\xi^R_{\qv,\sigma}=\delta_{\sigma\sigma'}$. 

The relation between the phonon Berry curvature and the intrinsic thermal Hall conductivity has been obtained in Ref.~\cite{Shi2012}. $\thc_{jk}$ in terms of $\varOmega^i_{\qv,\sigma}$ and $\upomega_{\qv,\sigma}$ is
\begin{align}
    \frac{\thc_{jk}}{T}=-\frac{1}{T^2}\int_0^\infty \diff E\, E^2 \sigma_{jk}(E) \frac{\diff n_B^{\text{eq}}(E)}{\diff E}=\int_0^\infty\diff x x^2 \frac{e^x}{(e^x-1)^2} \sigma_{jk}(x T)
    \label{appeq:kappa}
\end{align}
where 
\begin{align}
    \sigma_{jk}(E)=-\sum_\sigma \int \frac{\diff^3 q}{(2\pi)^3}\epsilon_{ijk}\varOmega^i_{\qv,\sigma}\Theta(E-\upomega_{\qv,\sigma})=-\frac{1}{4\pi^2}\int_0^\epsilon \diff E'\, E'^2 \sum_\sigma \int \frac{\diff\phi_\qv}{2\pi} \diff \theta_\qv \sin\theta_\qv \frac{1}{v_{\sigma}^3}\varOmega^i_{\qv,\sigma} \Theta(E-E').
    \label{appeq:sigma}
\end{align}
$n_B^{\text{eq}}(E)=1/(e^{\beta E}-1)$ is the equilibrium Bose distribution function. $\Sigma_\sigma$ sums over all eigenstates with positive energy. $\theta_{\ve{q}}, \phi_\qv$ are the polar coordinates for $\qv$, defined through $\ve{q}=\{q \sin\theta_\qv \cos \phi_\qv, q \sin \theta_\qv \sin \phi_\qv, q \cos \theta_\qv\}$.

Importantly, the behavior of $\kappa^H/T$ v.s.\ temperature is
determined by the functional form of $\sigma_{jk}(E)$ as we analyze
below. As an example, we compute $\kappa^H_{xy}(T)/T$ when the field
is $\ve{h}=h\bm{\hat{z}}$ perturbatively in the Hall viscosity
coefficient, keeping the first order in $\eta^H$. The antisymmetric $\mathsf{A}$ matrices for $\eta^H_{B_1 B_2}$, $\eta^H_{E_x E_y}$ are presented in Eq.~\eqref{eq:SGdef}.

\begin{figure}[t]
    \centering
    \includegraphics[width=0.4\columnwidth]{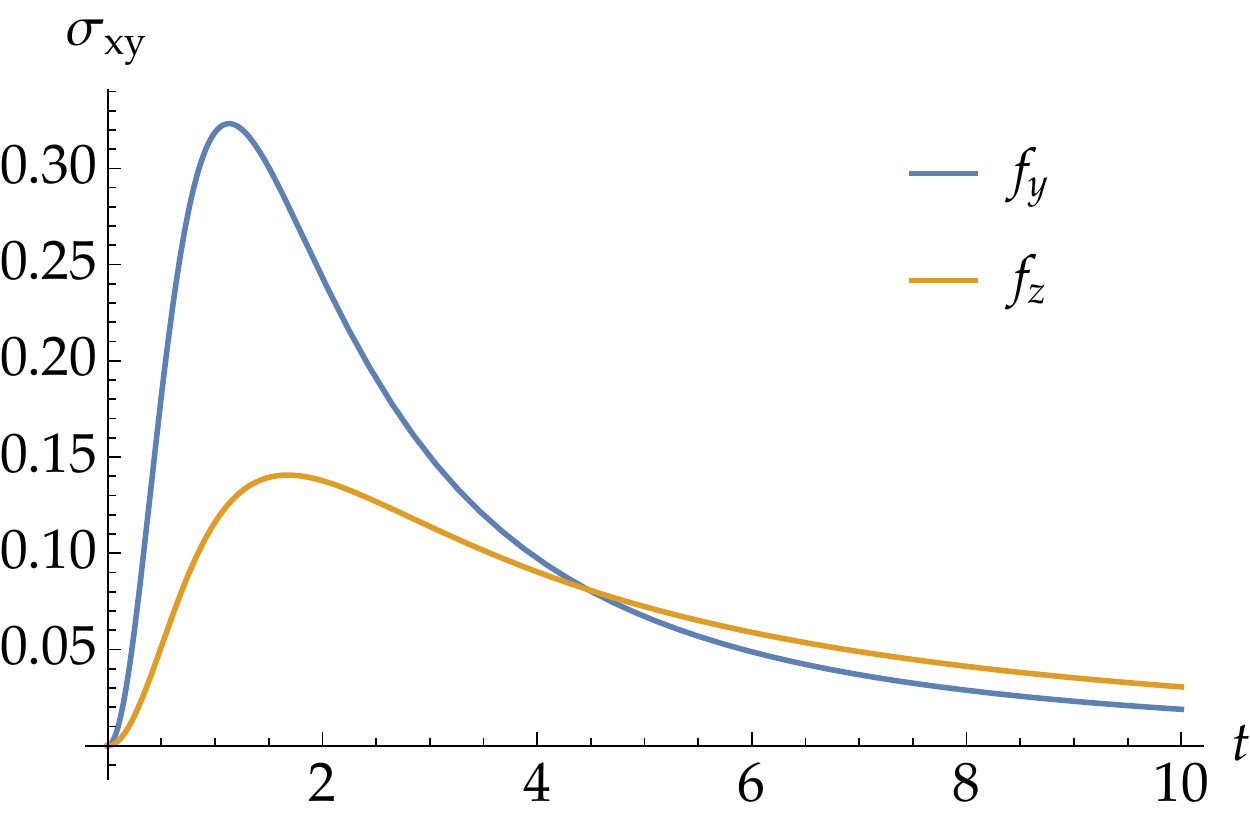}
    \caption{Behavior of $f_{y,z}(t)$ defined in Eq.~\eqref{appeq:f}. }
    \label{fig:f}
\end{figure}

To linear order in $\eta^H$, the contributions from different Hall viscosity terms simply add up. We find the Berry curvature $\varOmega^{z,\GG}_{\ve{q},\sigma}$ from $\eta^H_{\GG}$ of the two degenerate transverse phonon bands $\sigma=T_1,T_2$ and one longitudinal phonon band $\sigma=L$ as
\begin{align}
\varOmega^{z,B_1 B_2}_{\ve{q},T_1}+\varOmega^{z,B_1 B_2}_{\ve{q},T_2}=\frac{\gamma_{B_1B_2}v_T^2 \sin ^2 \theta_{\ve{q}} \left(v_L^2+3 v_T^2\right) \left[-4 \cos 2\theta_{\ve{q}} \left(3 \delta_y ^2 v_T^2+(v_T q) ^2\right)-20 \delta_y ^2 v_T^2+3 (v_T q) ^2 \cos 4 \theta_{\ve{q}}+(v_T q) ^2\right]}{2 \rho q  v_T  (v_T-v_L) (v_L+v_T) \left(2 \delta_y ^2 v_L^2-(v_T q) ^2 \cos 2\theta_{\ve{q}}+(v_T q) ^2\right)^2}\non\\
\varOmega^{z,B_1 B_2}_{\ve{q},L}=\frac{\gamma_{B_1B_2}v_L^2 \sin ^2 \theta_{\ve{q}} \left(3 v_L^2+v_T^2\right) \left[-4 \cos 2\theta_{\ve{q}} \left(3 \delta_y ^2 v_L^2+(v_L q) ^2\right)-20 \delta_y ^2 v_L^2+3 (v_L q) ^2 \cos 4\theta_{\ve{q}}+(v_L q) ^2\right]}{2\rho  v_L q  (v_L-v_T) (v_L+v_T) \left(2 \delta_z ^2 v_L^2-(v_L q) ^2 \cos 2\theta_{\ve{q}}+(v_L q) ^2\right)^2}\non\\
\varOmega^{z,E_x E_y}_{\ve{q},T_1}+\varOmega^{z,E_x E_y}_{\ve{q},T_2}=\frac{\gamma_{E_x E_y}v_T^2 \cos ^2\theta_{\ve{q}}\left(v_L^2+3 v_T^2\right) \left[-12 \cos 2\theta_{\ve{q}} \left(\delta_z ^2 v_T^2+(v_T q) ^2\right)-4 \delta_z ^2 v_T^2+3 (v_T q) ^2 \cos 4\theta_{\ve{q}}+9 (v_T q) ^2\right]}{\rho  v_T q  (v_T-v_L) (v_L+v_T) \left(2 \delta_z ^2 v_T^2-(v_T q) ^2 \cos2\theta_{\ve{q}}+(v_T q) ^2\right)^2}\non\\
\varOmega^{z,E_x E_y}_{\ve{q},L}=\frac{\gamma_{E_x E_y} v_L^2 \cos ^2\theta_{\ve{q}} \left(3 v_L^2+v_T^2\right) \left[-12 \cos 2\theta_{\ve{q}} \left(\delta_z ^2 v_L^2+(v_L q) ^2\right)-4 \delta_z ^2 v_L^2+3 (v_L q) ^2 \cos 4\theta_{\ve{q}}+9 (v_L q) ^2\right]}{\rho  v_L q  (v_L-v_T) (v_L+v_T) \left(2 \delta_z ^2 v_L^2-(v_L q) ^2 \cos 2\theta_{\ve{q}}+(v_L q) ^2\right)^2},
\label{appeq:Omega}
\end{align}
where $q=|\ve{q}|$.
From Eqs.~\eqref{appeq:Omega} and~\eqref{appeq:sigma}, we find 
\begin{align}
\sigma_{xy}^{B_1 B_2}(v_{\rm ph}\delta_y\, t)&=-\frac{4 \gamma_{B_1 B_2}}{v_{\rm ph} \rho } \left(-\frac{1}{ 1+t ^2}-\frac{ \left(3 +4 t ^2\right) \log \left(\sqrt{1+t ^2}+t \right)}{t ^3 \left(1+t ^2\right)^{3/2}}+\frac{3}{t ^2}\right)=-\frac{4 \gamma_{B_1 B_2}}{v_{\rm ph} \rho } f_y(t)\non\\
\sigma_{xy}^{E_x E_y}(v_{\rm ph}\delta_z \, t)&=-\frac{8 \gamma_{E_x E_y}}{v_{\rm ph} \rho } \left(\frac{ \left(3 +2 t ^2\right) \log \left(\sqrt{1+t ^2}+t \right)}{t ^3 \left(1+t ^2\right)^{3/2}}-\frac{3}{t ^2}\right)=-\frac{8 \gamma_{E_x E_y}}{v_{\rm ph} \rho } f_z(t).
\label{appeq:f}
\end{align}
Corrections at order $(v_L-v_T)/v_L$ are subleading and nonsingular,
and so they are ignored in the above expressions. $v_{\rm ph}$ denotes the averaged sound velocity. Note that $\sigma_{xy}^{\Gamma\Gamma'}(E)$ is only a function of $t=E/(v_{\rm ph}\delta_{\alpha(\Gamma\Gamma')})$, where $\alpha(B_1,B_2)=y, \alpha(E_x,E_y)=z$. For both $\sigma_{xy}^{B_1 B_2}$, $\sigma_{xy}^{E_x E_y}$, the function in the parentheses, defined as $f_{y,z}(t)$, increases as $\sim t^2$ for $t\ll 1$, and decreases to zero as $\sim \frac{1}{t^2}$ for $f_y(t)$, and as $\sim \frac{\ln t}{t^2}$ for $f_z(t)$ at $t\gg1$ (see Fig.~\ref{fig:f}).

From Eq.~\eqref{appeq:kappa} and~\eqref{appeq:f}, defining a
dimensionless temperature $\tilde{T}=T/(v_{\rm ph}\delta_\alpha)$,
\begin{align}
    \frac{\kappa_{xy}^{\Gamma\Gamma'}}{T}\sim F_{\alpha(\Gamma\Gamma')}\left(
  \tilde T\right), \qquad F_{\alpha(\Gamma\Gamma')}(\tilde T) = \int_0^\infty\diff x x^2 \frac{e^x}{(e^x-1)^2} f_{\alpha}(x\tilde T).
    \label{appeq:kappa2}
\end{align}
It is straightforward to see that the behavior of $\kappa^H/T$ is only a function of $\tilde T=T/(v_{\rm ph}\delta_\alpha)$. Also,
noting that the Boltzmann distribution decays exponentially at high temperature, $\frac{\kappa_{xy}^{\Gamma\Gamma'}}{T}$ at $\tilde T \ll 1$($T\ll v_{\rm ph} \delta_\alpha$) and $\tilde T \gg 1$($T\gg
v_{\rm ph}\delta_\alpha$) may be inferred from $f_\alpha(t)$ analytically. When $\tilde T \ll 1$, it mainly comes from $f_\alpha(t)\sim t^2$ at $t\ll 1$,
so we have $\frac{\kappa_{xy}^{\Gamma\Gamma'}}{T}\sim T^2$. When $\tilde T \gg 1$, the
scaling of $\frac{\kappa_{xy}^{\Gamma\Gamma'}}{T}$ is more complicated as it depends on
$f_\alpha(t)$ in the whole range. From the decaying behavior of
$f_\alpha(t)$ when $t\gtrsim 1$, we find that $\frac{\kappa_{xy}^{\Gamma\Gamma'}}{T}\sim
1/T^\zeta$ at $T\gg v_{\rm ph} \delta_\alpha$, with $\zeta\sim 1$ up to a logarithmic factor. 
In the intermediate regime, $\frac{\kappa_{xy}^{\Gamma\Gamma'}}{T}$
can be fit to an exponential empirically as $e^{-T/T_0^{\GG}}+\text{constant}$. From the numerical fit, we find $T_0^{B_1 B_2}\sim 2.3 v_{\rm ph} \delta_y$, $T_0^{E_x E_y}\sim 3.3 v_{\rm ph} \delta_z$.

\section{Contribution to $\eta^H$ from four-point correlation functions}
\label{app:twomagnon}

In this section, we discuss contributions to $\eta^H$ from four-point correlation functions of the spin fields. We argue that contributions at this order are subleading at low temperature, and can thus be ignored.

We consider the contribution to $\eta^H_{B_1 B_2}$ ($\eta^H_{E_x E_y}$
is similar).  
From Eq.~\eqref{eq:1} and Tab.~\ref{tab:myops}, we find
\begin{align}
&\langle \mac{O}_{B_1}(-\qv,\tau) \mac{O}_{B_2}(\qv,0) \rangle_{\omega_n}\non\\
\sim &  \frac{1}{L^2} \sum_{\kv,\kv'} \left(\varsigma_1 \langle  n_{y,-\bm{k}-\qv_\pp}(\tau)n_{y,\bm{k}}(\tau)n_{y,\bm{k}'+\qv_\pp}(0){\rm m}_{z,-\bm{k}'}(0)\rangle_{\omega_n}+\varsigma_2 \langle n_{z,-\bm{k}-\qv_\pp}(\tau)n_{z,\bm{k}}(\tau)n_{z,\bm{k}'+\qv_\pp}(0){\rm m}_{y,-\bm{k}'}(0)\rangle_{\omega_n}\right)\non\\
\sim& T\frac{1}{L^2} \sum_{\kv,m} \left( \varsigma_1 \langle  n_{y,-\bm{k}-\qv_\pp}{\rm m}_{z,\bm{k}+\qv_\pp}\rangle_{\omega_m}  \langle n_{y,\bm{k}}(\tau)n_{y,-\bm{k}}\rangle_{-\omega_m+\omega_n} +\varsigma_2 \langle n_{z,-\bm{k}-\qv_\pp}{\rm m}_{y,\bm{k}+\qv_\pp}\rangle_{\omega_m}  \langle  n_{z,\bm{k}}(\tau)n_{z,-\bm{k}}\rangle_{-\omega_m+\omega_n}\right).
\label{appeq:etabubble}
\end{align}
where $\varsigma_1, \varsigma_2$ are dimensionless coefficients that determine the magnitude and sign of the two types of four-point correlations. Following the analysis in Sec.~\ref{sec:norm-oper-four}, an overall factor $\frac{\lambda_\Gamma \lambda_{\Gamma'}}{d_z} \left(\frac{S^2}{a_0^2}\right)^2 \chi h$ is absorbed into $\sim$.

For simplicity, we consider the limit when $q=|\ve{q}|\rightarrow 0$. After
summing over the Matsubara frequencies $\omega_m$ and analytical continuing to real frequencies, $\eta^H_{B_1 B_2}$ from Eq.~\eqref{appeq:etabubble} can be expressed as
\begin{align}
    \left[\eta_{B_1,B_2}^{H}(\qv)\right]^{ (2)}\sim&\frac{\lambda_\Gamma \lambda_{\Gamma'}}{d_z} \left(\frac{S^2}{a_0^2}\right)^2 \chi h \,\frac{a_0^2}{S} \, v_m g a_0 \non\\
   & \frac{1}{a_0^2} \frac{1}{N_x N_y}\sum_\kv \left\{\frac{\varsigma_1}{(\upomega_{y,\kv}+\upomega_{y,\kv-\qv_\pp})^2}\left(\frac{1}{\upomega_{y,\kv}}\coth{\frac{\beta \upomega_{y,\kv}}{2}}+\frac{1}{\upomega_{y,\kv-\qv_\pp}}\coth{\frac{\beta \upomega_{y,\kv-\qv_\pp}}{2}}\right)\right.\non\\
    &\qquad\qquad\quad\,\,\left.-\frac{\varsigma_2}{(\upomega_{z,\kv}+\upomega_{z,\kv-\qv_\pp})^2}\left(\frac{1}{\upomega_{z,\kv}}\coth{\frac{\beta \upomega_{z,\kv}}{2}}+\frac{1}{\upomega_{z,\kv-\qv_\pp}}\coth{\frac{\beta \upomega_{z,\kv-\qv_\pp}}{2}}\right)\right\}\non\\
    \xrightarrow{q\rightarrow 0}&\gamma_{B_1 B_2} v_m^3 g a_0\frac{1}{a_0^2}\int_{-\frac{\pi}{a_0}}^{\frac{\pi}{a_0}}\int_{-\frac{\pi}{a_0}}^{\frac{\pi}{a_0}}\frac{\diff^2 k}{(2\pi/a_0)^2}\left\{\frac{\varsigma_1}{\upomega_{y,\kv}^3}\coth{\frac{\beta \upomega_{y,\kv}}{2}}
   -\frac{\varsigma_2}{\upomega_{z,\kv}^3}\coth{\frac{\beta \upomega_{z,\kv}}{2}}\right\}\non\\
    \sim &\gamma_{B_1 B_2} v_m^3 g a_0 \frac{1}{T v_m^2}\left[\varsigma_1 \mac{F}\left(\frac{v_m \delta_y}{2T}\right)-\varsigma_2 \mac{F}\left(\frac{v_m \delta_z }{2T}\right)\right]\non\\
    \sim&\gamma_{B_1 B_2} v_m g a_0  
    \begin{cases}
    \frac{1}{v_m \delta_y} & T\ll v_m \delta_y\\
    \frac{T}{(v_m \delta_y)^2 } & T\gg v_m \delta_y
    \end{cases}
    \label{eq:eta1br}
\end{align}
where $\mac{F}(x)=\int_x^\infty \diff z \frac{1}{z^2}\coth z$, and $
\delta_y\ll  \delta_z$ has been applied to obtain the last line. As our interest here is the relative strength of four-point contribution compared to two-point contribution, numerical coefficients are subsumed in the ``$\sim$". The superscript $``(2)"$ denotes that it is a second order contribution. Note
that the contribution from 2 magnons is much smaller than that from 1 magnon by a factor $\frac{g T}{v_m a_0^{-1} }\sim \frac{T}{J S^2}\ll 1$. For this reason, only contributions from 1 magnon terms are considered further in the main text.

\section{Microscopic Hamiltonian}
\label{app:MicroscopicH}
In this section, we present the microscopic spin Hamiltonian and spin-lattice coupling Hamiltonian. The microscopic analysis is helpful to infer the spin gap and relevant spin-lattice coupling in the low energy effective action derived in the main text from a symmetry analysis. 

The microscopic Hamiltonian must be invariant under the grey magnetic space group $\mathsf{G}$ (see Table~\ref{tab:MagneticSG} ) of
the crystal.  It is useful to decompose the part involving the spin as:
\begin{align}
    H &=H_{\rm s}+H_{\rm sl}\non\\
    &=H_{\rm s}^{(0)}+\lambda_{SOC}H_{\rm s}^{(1)}+H_{\rm sl}^{(0)}+\lambda_{SOC}H_{\rm sl}^{(1)},
    \label{eq:Hmicroscopic}
\end{align}
where ${\rm s}$ refers to ``spin'' and ${\rm sl}$ to ``spin-lattice coupling''. We assume weak SOC (as e.g.\ appropriate for cuprates), and accordingly
separate terms of 0th order in SOC ($H_{\rm s}^{(0)}+H_{\rm sl}^{(0)}$) and those
that require SOC ($H_{\rm s}^{(1)}+H_{\rm sl}^{(1)}$).     This is defined by symmetry:  $H_{\rm s}^{(0)}+H_{\rm sl}^{(0)}$ also has a global $SO(3)$ spin-rotation symmetry, so is also invariant under $SO(3)_s$, while $H_{\rm s}^{(1)}+H_{\rm sl}^{(1)}$ must break $SO(3)_s$. 

\emph{$H_{\rm s}$} --The spin Hamiltonian reads
\begin{align}
    H_s^{(0)}&=\sum_{i} J (\ve{S}_\ver \cdot \ve{S}_{\ver \pm \hat{\ve{x}}}+\ve{S}_\ver \cdot \ve{S}_{\ver \pm \hat{\ve{y}}})\non\\
        H_s^{(1)}&= \sum_\ver J'_1 (S^x_\ver S^x_{\ver\pm \bm{\hat{x}}}-S^y_\ver S^y_{\ver\pm \bm{\hat{x}}}-S^x_\ver S^x_{\ver\pm \bm{\hat{y}}}+S^y_\ver S^y_{\ver\pm \bm{\hat{y}}})+J'_z (S^z_\ver S^z_{\ver+\bm{\hat{x}}}+S^z_\ver S^z_{\ver+\bm{\hat{y}}})\non\\
        &\quad+J'_2(S^x_\ver S^y_{\ver\pm \bm{\hat{a}}}+S^y_\ver S^x_{\ver\pm \bm{\hat{a}}}-S^x_\ver S^y_{\ver\pm \bm{\hat{b}}}-S^y_\ver S^x_{\ver\pm \bm{\hat{b}}} )
    \label{appeq:Hs1}
\end{align}  
where, as before, $\bm{\hat{a},\hat{b}}=\frac{1}{\sqrt{2}}(\pm
\bm{\hat{x}}+\bm{\hat{y}})$ are second neighbor vectors. The nearest neighbor exchange
dominates and $J \sim 0.1$~eV.  
From Refs.~\cite{Anisotropic1995,Benfatto2006}, the anisotropy gap $\Delta_{y,z}$ can be estimated for \scoc \, from the spin exchange interactions that require SOC. 
The XXZ anisotropic exchange $J'_z$ is positive, which stabilizes the staggered order in the
$xy$ plane. $J'_z$ introduces a gap for out-of-plane
fluctuations $\Delta_z=4SJ'_z\sim 3$~meV, where the factor of $4$ comes from
the coordination number of square lattice. While $J'_{1,2}\ll J$,
$J'_1$ is still important to introduce
an {\em{effective}} mass gap to the in-plane Goldstone mode at order
$1/S$. In Ref.~\cite{Anisotropic1995}, by studying the dependence of
the quantum zero point energy on the angle $\phi$ of the N\'eel order relative to $\vhat{x}$, it is shown that $J'_1$ stabilize N\'eel order with $\phi=\pi/4 \text{ mod } \pi/2$, and thus the in-plane Goldstone mode acquires an effective gap, which is estimated to be $\Delta_{y}=0.05$~meV when $S=1/2$.

\emph{$H_{\rm sl}$} --
In this work, we assume the spin-lattice coupling comes from the
modification to the exchange coupling when the direction and length of
the bond change due to an elastic deformation. The microscopic
spin-lattice Hamiltonian can be obtained by finding all the terms
composed of $\mathcal{E}_\Gamma$ and $\mathcal{O}_{\Gamma}$ invariant
under $\mathsf{G}$, where $\mathcal{E}_\Gamma$ is the strain tensor
decomposed into irreps of ${\rm D}_{4h}$ (see Eq.~\eqref{eq:strainT}
in the main text), and on the lattice we take $\mathcal{O}_{\Gamma}$
to be composed of bilinears of spin operators on nearby sites, chosen to transform under ${\rm D}_{4h}$ with the same irrep $\Gamma$. In real space, we find 
\begin{align}
    H_{\rm sl}^{(0)}=&\lambda_{A_1}\sum_{\ver}(\mathcal{E}_{A_1})_{\ver}(\ve{S}_{\ver}\cdot \ve{S}_{\ver \pm\vhat{x}}+\ve{S}_{\ver}\cdot \ve{S}_{\ver\pm\vhat{y}})+\frac{\lambda_{B1}}{2}\sum_{\ver}(\mathcal{E}_{B_1})_{\ver}\left(\ve{S}_{\ver}\cdot \ve{S}_{\ver+\bm{\hat{x}}}-\ve{S}_{\ver}\cdot \ve{S}_{\ver+\bm{\hat{y}}}\right)+\non\\
    &\frac{\lambda_{B2}}{2}\sum_{\ver}(\mathcal{E}_{B_2})_{\ver}\left(\ve{S}_{\ver}\cdot \ve{S}_{\ver+\vhat{a}}-\ve{S}_{\ver}\cdot \ve{S}_{\ver+\vhat{b}}\right).
    \label{appeq:Hsl0}
\end{align}
Here we denote by $(\mathcal{E}_{\Gamma})_{\bm{x}}$ the strain at
position $\bm{x}$ -- note that since the strain is slowly varying,
shifts of this coordinate by order one displacements do not modify the
results.  The XXZ anisotropy may be ignored here {\em in the
  spin-lattice coupling} as it is negligibly small and not essential
(it is, however, important in the spin Hamiltonian itself).
The anisotropic spin-lattice coupling terms, keeping spin operators
on the nearest possible pairs of spins in each channel, give $H_{\rm
  sl}^{(1)}$ in the form
\begin{align}
    H_{\rm sl}^{(1)}=
    &\frac{\lambda'_{B1}}{2}\sum_{\ver}(\mathcal{E}_{B_1})_{\ver}\left(S^x_{\ver} S^x_{\ver\pm {\bm{\hat{x}}}}-S^y_r S^y_{\ver\pm {\bm{\hat{y}}}}+S^x_{\ver} S^x_{\ver\pm {\bm{\hat{y}}}}-S^y_{\ver} S^y_{\ver\pm {\bm{\hat{x}}}}\right)+\non\\
    &\frac{\lambda'_{B2}}{2}\sum_{\ver}(\mathcal{E}_{B_2})_{\ver}\left(S^x_{\ver} S^y_{\ver\pm \vhat{a}}+S^y_{\ver} S^x_{\ver\pm \vhat{a}}+S^x_{\ver} S^y_{\ver\pm \vhat{b}}+S^y_{\ver} S^x_{\ver\pm \vhat{b}}\right)+\non\\
    &\frac{\lambda'_{E}}{2}\sum_{\ver} (\mathcal{E}_{E_x})_{\ver}\left[(1+\beta)\left(S^x_{\ver} S^z_{\ver\pm {\bm{\hat{x}}}}+S^z_{\ver} S^x_{{\ver}\pm {\bm{\hat{x}}}}\right)+(1-\beta)\left(S^x_{\ver} S^z_{{\ver}\pm {\bm{\hat{y}}}}+S^z_{\ver} S^x_{{\ver}\pm {\bm{\hat{y}}}}\right)\right]+\non\\
    &\frac{\lambda'_{E}}{2}\sum_{\ver} (\mathcal{E}_{E_y})_{\ver}\left[(1-\beta)\left(S^y_{\ver} S^z_{{\ver}\pm {\bm{\hat{x}}}}+S^z_{\ver} S^y_{{\ver}\pm {\bm{\hat{x}}}}\right)+(1+\beta)\left(S^y_{\ver} S^z_{{\ver}\pm {\bm{\hat{y}}}}+S^z_{\ver} S^y_{{\ver}\pm {\bm{\hat{y}}}}\right)\right].
    \label{appeq:Hsl1}
\end{align}
Note that in the $\Gamma=E$ channel, there are two symmetry allowed
spin-lattice coupling terms, of strength $\lambda'_E$ and $\beta
\lambda'_E$, where $\beta$ is a dimensionless parameter defined by
Eq.~\eqref{appeq:Hsl1}.

The above microscopic spin-lattice couplings give rise to continuum
spin-lattice couplings of the form in Eq.~\eqref{eq:Lsl} of the main
text by taking a continuum limit.  In particular, one uses the NLSM decomposition with $\bm{S}_{\bm{r}} =S
\bm{e}_{\bm{r}}$ with $\bm{e}_{\bm{r}}$ in Eq.~\eqref{appeq:NLSMN},
and expresses the Hamiltonian thereby in terms of $\bm{n}$ and
$\bm{m}$, which are presumed to be slowly varying functions of
position.   To zeroth order in the gradient expansion of these fields,
derivatives $\partial_\mu \bm{n},\partial_\mu\bm{m}$ are neglected.
The result then takes the {\em schematic} form
\begin{align}
    H_{\rm sl}^{(loc)}\sim \sum_z \frac{1}{a_0^2}\int \diff x \diff y
     &\left[\lambda_{A_1}\mathcal{E}_{A_1}\bm{s}\cdot \bm{s}+\lambda'_{B1}\mathcal{E}_{B_1}\left(s^xs^x-s^y s^y\right)+\lambda'_{B2} \mathcal{E}_{B_2}\left(s^x s^y+s^y s^x\right)+\right.\non\\
     &\left. \lambda'_{E} \mathcal{E}_{E_x}\left(s^x s^z+s^z s^x\right)+\lambda'_{E}\mathcal{E}_{E_y}\left(s^y s^z+s^z s^y\right)\right].
    \label{appeq:Hslloc}
\end{align}
Here each term represents a sum of two contributions, one with
$\bm{s}=S\bm{m}$, and another with $\bm{s}=S\bm{n}$, and some sign
differences may appear between these terms.  
Here, we summed over 2d magnetic layers ($\sum_z$), where the layer index is $z$. In the limit of decoupled magnetic layers, we considered the continuous limit as $\sum_\ver \rightarrow \sum_z \frac{1}{a_0^2} \int \diff x \diff y$ (see the discussion of Eq.~\eqref{eq:7} for more details).  From Eq.~\eqref{appeq:Hslloc}, we see that 
$H_{\rm sl}$ in all channels $\Gamma\neq A_1$ requires breaking
$SO(3)_s$, and thus requires SOC microscopically. Eq.~\eqref{appeq:Hslloc} give results
consistent with the general symmetry analysis in terms of continuous
spin fields $\ve{n}$ and $\bm{m}$ (c.f. Table~\ref{tab:myops}). 


 \section{Phonon Hall viscosity for in-plane magnetic field}
\label{app:etaHy}
For completeness, we have also analyzed the phonon Hall viscosity for
in-plane magnetic fields.  For concreteness, the field is applied along the $y$-axis. Such a scenario can
be relevant to the phonon Hall effect when the heat current is applied perpendicular
to the CuO$_2$ planes, i.e.\ upon studying $\kappa_{xz}$. The analysis follows that for the out-of-plane field presented in the main text. Here, we summarize the main results. 

The magnetic space group for a paramagnet and AFM with staggered order in the $xy$ plane at an
arbitrary $\phi\neq 0 \text{ mod } \pi/4$ and $\phi=0$ relative to $\vhat{x}$, when the field is along $\bm{\hat{y}}$, is summarized in Table~\ref{tab:MagneticSGy}.
\begin{table}[H]
\centering
\renewcommand{\arraystretch}{1.5}
\begin{tabular}{c|l|l|l}
\hline
&  \multirow{2}{3cm}{\quad\quad\quad\quad zero field} & \multicolumn{2}{c}{$\ve{h}=h\bm{\hat{y}}$}\\
\cline{3-4}
&   & \quad\quad lattice and spin & \quad lattice effective \\
\hline
 paramagnet & \;$\mathsf{G}=P4/mmm1'$\; &
                                          \;$\mathsf{G}(\bm{0},h\bm{\hat{y}})=\langle X, Y, C_{2y}, i, \trl {C_{2z}} \rangle$  &  \;$\mathsf{G}^{\rm eff}(\bm{0},h\bm{\hat{y}})=\langle C_{2y}, i, \trl {C_{2z}} \rangle$ \\
  \hline
  high sym.\ AFM & \;$\mathsf{G}(\bm{\hat{x}},\bm{0})=\langle
                   i,\trl X,\trl Y, C_{2x},  \trl C_{2z}\rangle$\;
                                                      &\;$\mathsf{G}(\bm{\hat{x}},h\bm{\hat{y}})=\langle  i,XY, \trl C_{2z} \rangle$ &
                                                                   \;$\mathsf{G}^{\rm eff}(\bm{\hat{x}},h\bm{\hat{y}})=\langle  i , \trl C_{2z} \rangle$ \\
\hline
 low sym.\ AFM & \;$\mathsf{G}(\bm{\hat{e}},\bm{0})=\langle
                 i,\trl X,\trl Y, \trl C_{2z}\rangle$\; &
                                                         \;$\mathsf{G}(\bm{\hat{e}},h\bm{\hat{y}})=\langle i, XY,  \trl C_{2z} \rangle$ & \;$\mathsf{G}^{\rm eff}(\bm{\hat{e}},h\bm{\hat{y}})=\langle i, \trl C_{2z}\rangle$ \\
\hline
\end{tabular}
\caption{\label{tab:MagneticSGy} Summary of the magnetic space group (in the
  Hermann-Mauguin notation) for layer group
  $G=P4/mmm$ with $\ve{h}=h\vhat{y}$. Same to Tab.~\ref{tab:MagneticSG} in the main text, rows denote different values for $\bm{n}$: without
  magnetic order, and for the two distinct orientations of the
  staggered magnetization in the $xy$ plane (here $\bm{\hat{e}}$ is a
  generic vector {\em not} along a high-symmetry axis in the plane), and
  columns specify the zero and finite field cases, and the effective
  magnetic group for the effective lattice theory. The symbol $\langle \cdot \rangle$ indicates
  the group generated by the ``$\cdot$'' operations.
}
\end{table}
  
Note that $\eta^H_{E_x E_y}=0$ due to the $\trl {C_{2z}}$ symetry, so
there is no AFE. To generate a nonzero phonon Berry curvature $\varOmega^y$,
both time-reversal and mirror symmetries perpendicular to the $xz$
plane ($\sigma_{h}$ and $\sigma_{vx}$ in the ${\rm D}_{4h}$ group) should be
broken. The symmetry-allowed Hall viscosity that generates a nonzero phonon Berry curvature reads 
\begin{align}
    \mathcal{S}_{{\rm PHV},h \vhat{y}}=\int\diff^3 x\diff \tau \, \{ 
    \eta^H_{A_1 E_x} (\mathcal{E}_{A_1}\dot{\mathcal{E}}_{E_x}-\dot{\mathcal{E}}_{A_1}\mathcal{E}_{E_x})+
    \eta^H_{B_1 E_x} (\mathcal{E}_{B_1}\dot{\mathcal{E}}_{E_x}-\dot{\mathcal{E}}_{B_1}\mathcal{E}_{E_x})+
    \eta^H_{B_2 E_y} (\mathcal{E}_{B_2}\dot{\mathcal{E}}_{E_y}-\dot{\mathcal{E}}_{B_2}\mathcal{E}_{E_y})\}. 
    \label{eq:HallParaHx}
\end{align}
To obtain the Hall viscosity induced by spin-lattice coupling, we find
$\mac{O}_\Gamma(\ve{x},\tau)$ (see Eq.~\eqref{eq:Lsl}) in terms of the
spin fields as tabulated in Table~\ref{tab:Ohy}.
\begin{table}[h]
\begin{tabular}{c|| c c c c}
    & $\epsilon\, n$ & $\epsilon\, n n $ & $\epsilon\, m m $ & $\epsilon\, n m m$\\
    \hline
    $\mac{O}_{A_1}$  & & $n_y n_y, n_z n_z$ & $h\, {\rm m}_y $ & \\
    $\mac{O}_{B_1}$  & & $n_y n_y, n_z n_z$ & $h\, {\rm m}_y $ & \\
    $\mac{O}_{B_2}$  & $n_y$ &   & $-h\, n_y {\rm m}_y$   & $h\, n_y {\rm m}_y , h\,n_z  {\rm m}_z $ \\
    $\mac{O}_{E_x}$  & $n_z$ &   & $-h\, n_y {\rm m}_z$  & $h\, n_z {\rm m}_y , h\,n_y  {\rm m}_z $ \\
    $\mac{O}_{E_y}$  & & $n_y n_z$ &  $h\, {\rm m}_z $ & \\
    \end{tabular}.
\caption{Magnetic operators in the high symmetry AFM in the
      presence of a small field along the $y$-axis. Other factors, e.g. $S^2,\chi, 1/a_0^2,n_0$, have been omitted in the table.}
      \label{tab:Ohy}
     \end{table}

At leading order in $1/S$, this gives
\begin{align}
    \eta_{A_1,E_x}^H\sim h \langle  {\rm m}_y n_z \rangle,\quad
    \eta_{B_1,E_x}^H\sim h \langle  {\rm m}_y n_z \rangle,\quad
    \eta_{B_2,E_y}^H\sim h \langle  {\rm m}_z n_y \rangle.
\end{align}
Note that at leading order in $1/S$, the Hall viscosity coefficients
are determined by the same sets of spin correlators as when $\ve{h}=h
\bm{\hat{z}}$. However, this is restricted to when the in-plane
field is perpendicular to the staggered order. If the external field
also has components parallel to the staggered order, we find that
$\langle n_y n_z\rangle_\omega \sim ih\omega$ also contribute to a nonzero Hall viscosity.

\end{document}